\documentclass[12pt, twoside]{article}\usepackage[]{graphicx}\usepackage[]{color}
\makeatletter
\def\maxwidth{ %
  \ifdim\Gin@nat@width>\linewidth
    \linewidth
  \else
    \Gin@nat@width
  \fi
}
\makeatother

\definecolor{fgcolor}{rgb}{0.345, 0.345, 0.345}
\newcommand{\hlkwb}[1]{\textcolor[rgb]{0.69,0.353,0.396}{#1}}%

\usepackage{framed}
\makeatletter
 {\par\unskip\endMakeFramed%
 \at@end@of@kframe}
\makeatother

\definecolor{shadecolor}{rgb}{.97, .97, .97}
\definecolor{messagecolor}{rgb}{0, 0, 0}
\definecolor{warningcolor}{rgb}{1, 0, 1}
\definecolor{errorcolor}{rgb}{1, 0, 0}
\newenvironment{knitrout}{}{} % an empty environment to be redefined in TeX

\usepackage{alltt}

\usepackage[margin=1in]{geometry}

\usepackage[font=doublespacing]{caption}
\usepackage{setspace}

\usepackage{amsthm, amssymb, bm, booktabs, dsfont}
\usepackage[leqno]{amsmath}

\usepackage{xparse}

\RequirePackage[colorlinks,linkcolor=blue,citecolor=blue,urlcolor=blue]{hyperref}
\usepackage{xr}
\usepackage[natbibapa, nodoi]{apacite}
\usepackage{etoolbox}

\AtBeginEnvironment{APACrefURL}{\renewcommand{\url}[1]{}}
\usepackage{cleveref}
\usepackage{placeins}
\usepackage[nolists, nomarkers, noheads, tablesonly]{endfloat}

\def\equationautorefname~#1\null{%
  (#1)\null
}

  \renewcommand{\vec}[1]{\boldsymbol{#1}}
  \renewcommand{\v}[1]{\vec{#1}}

  \DeclareMathOperator{\diag}{diag}
  \DeclareMathOperator{\HS}{HS}
  \DeclareMathOperator{\Cor}{Cor}

  \newcommand{\card}[1]{\left\vert#1\right\vert}
  \newcommand{\paren}[1]{\left(#1\right)}
  \newcommand{\brc}[1]{\left\{#1\right\}}
  \newcommand{\brk}[1]{\left[#1\right]}
  \newcommand{\inv}[1]{#1^{-1}}
  \newcommand{\given}[2]{\left.#1\right\vert#2}

  \newcommand{\E}[1]{E\left[#1\right]}

  \newcommand{\Odds}[1]{\text{Odds}\left(#1\right)}

  \NewDocumentCommand\distn{mg}{
    #1\IfNoValueTF{#2}{}{\left(#2\right)}
  }

  \newcommand{\ns}{{n_s}}
  \newcommand{\nr}{{n_r}}
  \newcommand{\nt}{{n_t}}

  \newcommand{\vs}{{\v s}}
  \newcommand{\vr}{{\v r}}
  \newcommand{\vu}{{\v u}}
  \newcommand{\vv}{{\v v}}
  \newcommand{\Dz}{{\mathcal D_Z}}
  \newcommand{\Dy}{{\mathcal D_Y}}

  \newcommand*{\nolink}[1]{%
    \begin{NoHyper}#1\end{NoHyper}%
  }

\title{Remote effects spatial process models for modeling teleconnections}

\usepackage{fancyhdr}
\usepackage{authblk}
\author[1]{Joshua Hewitt}
\author[1]{Jennifer A. Hoeting}
\author[2]{James M. Done}
\author[2]{Erin Towler}
\affil[1]{Colorado State University}
\affil[2]{National Center for Atmospheric Research}

\date{\vspace{-5ex}}

\IfFileExists{upquote.sty}{\usepackage{upquote}}{}
\begin{document}

\singlespace

\maketitle

\doublespace

\begin{abstract}

{\textbf{Abstract:}}  While most spatial data can be modeled with the assumption
that distant points are uncorrelated, some problems require dependence at both
far and short distances.  We introduce a model to directly incorporate
dependence in phenomena that influence a distant response.  Spatial climate 
problems often have such modeling needs as data are influenced by local factors 
in addition to remote phenomena, known as teleconnections.  Teleconnections 
arise from complex interactions between the atmosphere and ocean, of which the 
El Ni\~{n}o--Southern Oscillation teleconnection is a well-known example.
Our model extends the standard geostatistical modeling framework to account for 
effects of covariates observed on a spatially remote domain.  We frame our 
model as an extension of spatially varying coefficient models.  Connections to
existing methods are highlighted and further modeling needs are addressed by
additionally drawing on spatial basis functions and predictive processes. 
Notably, our approach allows users to model teleconnected data without 
pre-specifying teleconnection indices, which other methods often require.  
We adopt a hierarchical Bayesian framework to conduct inference and make 
predictions. The method is demonstrated by predicting precipitation in Colorado 
while accounting for local factors and teleconnection effects with Pacific 
Ocean sea surface temperatures. We show how the proposed model improves upon 
standard methods for estimating teleconnection effects and discuss its utility 
for climate  applications.
\end{abstract}


\section{Introduction}
\label{sec:intro}

While most spatial data can be modeled with the assumption that distant points
are uncorrelated, some problems require dependence at both far and short
distances. Spatial climate data is an example of the latter, as it is influenced
by local (i.e., short distance) factors, as well as by remote 
(i.e., far distance) phenomena called teleconnections. Teleconnections refer to 
changes in patterns of large-scale atmospheric circulation that can drive 
changes in temperature and precipitation in distant regions 
\citep[e.g.,][]{Tsonis2008, Ward2014}. Most teleconnection modeling approaches 
in the statistical literature do not explicitly estimate dependence within 
remote phenomena.  The statistical literature includes spatially varying 
coefficient models, analogs, and covariance matrix estimation 
\citep{Calder2008, Wikle2003, McDermott2016, Choi2015}.  Explicitly modeling
dependence in remote phenomena can add physically sensible structure that 
improves prediction accuracy and addresses some modeling challenges.
We propose a geostatistical model that addresses this unmet modeling need for
teleconnection.

Teleconnections can be forced by changes in sea surface temperature (SST), and 
there have been many observational and modeling studies studying the link 
between SSTs, circulation patterns, and impacts on global and regional climate. 
Several seminal studies connect U.S. precipitation with SST anomalies in the 
tropical Pacific due to the El Ni\~{n}o--Southern Oscillation teleconnection 
(ENSO) \citep{Montroy1997, Montroy1998}, as well as with SST anomalies in the 
Pacific \citep[e.g.,][]{Dong2015}. The ENSO teleconnection has been critical
in seasonal climate forecasting \citep{Goddard2001}, and decadal variability of 
sea surface temperature anomalies have been identified as a source of potential 
skill for decadal predictions that look out one year to a decade 
\citep{Meehl2009}. In terms of the latter, decadal predictions produced from 
global climate models (GCMs) have shown skill in reproducing ocean and land 
temperatures, and less skill in precipitation \citep{Meehl2014a}. This is the 
general finding for GCMs: while GCMs perform poorly in predicting precipitation 
directly, they can skillfully reproduce surface temperatures and large-scale 
patterns \citep{Flato2013}.  Direct precipitation prediction by GCMs is 
challenging because of complex and interacting multi-scale physical 
precipitation processes, resulting in large uncertainty in future precipitation 
patterns \citep{Deser2012}. As such, this 
provides a motivating example for demonstrating a teleconnection model that can 
be used in conjunction with GCM output to estimate impacts on precipitation.

Developing a teleconnection model for application with GCM output has overlaps 
with the burgeoning field of statistical downscaling. Statistical downscaling 
methods use large-scale variables to draw inference on regional variables. 
Similar to what is being proposed here, a type of statistical downscaling 
called perfect prognosis downscaling  \citep{Maraun2010} develops a statistical 
relationship between observed large-scale predictors and local-scale weather 
phenomena \citep[e.g.,][]{Wilby1998, Bruyere2012, Towler2016}.  Common models
used for perfect prognosis downscaling do not explicitly model spatial 
dependence.  \citet{Maraun2010} review methods used in the climate literature, 
which include linear models, analogs, and machine learning techniques like 
neural networks.  Dependence is often indirectly modeled by using principle 
component or canonical correlation basis functions as predictors and applying 
various corrections to uncertainties \citep[cf.][]{Karl1990}. After
statistical relationships are developed and validated on observed datasets,
models can be applied to large-scale GCM output to obtain an estimate of the 
desired predictant. Clearly, perfect prognosis methods are highly dependent on 
the selected predictors and model \citep{Fowler2007}. 

We propose a remote effects spatial process (RESP) model that extends spatially
varying coefficient models to directly model dependence in remote phenomena and
address several modeling challenges.  Spatially modeling dependence in remote 
phenomena adds sensible structure to teleconnection models which, in turn,  
allows better use of the data than standard models.  
Standard spatially varying coefficient models regress a 
local response $Y\paren{\vs, t}$ with spatio-temporal error $w\paren{\vs, t}$ 
onto local covariates $\v x\paren{\vs, t}$ through
\begin{align}
\label{eq:soa_model}
    Y\paren{\vs, t} = \v x\paren{\vs, t}^T \v\beta +
      \v z\paren{t}^T \v\theta\paren\vs + w\paren{\vs, t}
\end{align}
which includes adjustment for spatially-varying effects
$\v\theta\paren\vs\in\mathbb R^k$ associated with a second vector
$\v z\paren{t}\in\mathbb R^k$ of $k$ covariates
\citep[Section 9.6.2]{Banerjee2015}.  As applied to teleconnection, the
covariate vector $\v z\paren{t}$ contains one or more indices that quantify the
overall strength or state of large-scale patterns, like ENSO or the
North Atlantic Oscillation \citep{Calder2008, Wikle2003}.  While effective,
the model \autoref{eq:soa_model} assumes relevant large-scale patterns are
known a priori (e.g., ENSO).  However, relevant teleconnection indices can
depend on the study region and thus be unknown at the start of an analysis 
\citep{Towler2016}.  The spatially varying coefficient model 
\autoref{eq:soa_model} will be inefficient if driven by poorly chosen 
teleconnection indices. Standard formulations of \autoref{eq:soa_model} also
model within-site covariances for spatially varying effects
$\Lambda = \textrm{Cov}\paren{\v\theta\paren\vs} \in\mathbb R^{k\times k}$
with non-spatial covariance matrices.  While the issue may be less important for
orthogonal teleconnection indices, typical indices are defined with respect to
different covariates and zonal averages so may not be orthogonal
\citep[cf.][]{Ashok2007, Mantua1997}. Instead, teleconnection indices may have
spatial structure induced by remote covariates.  The RESP model introduced below
directly incorporates remote covariates instead of using teleconnection indices
and can offer potential improvement for the a priori and spatial structure
concerns (\Cref{sec:model}). Notably, the RESP model does not lose generality
since direct connections can be drawn to standard spatially varying coefficient
models (\Cref{sec:reparam}).

More generally, the RESP model represents a less-common class of spatial
analysis problems that provide rich opportunities for study. 
We introduce our teleconnection model in the general context of a spatial 
regression problem involving local and spatially remote covariates 
(\Cref{sec:model}).  The local and spatially remote covariates are allowed to 
have different spatial correlations structures reflecting their different 
relationships with the response.  
\autoref{fig:tele_schematic} schematically illustrates the general
teleconnection problem in which local $\v x\paren{\v s, t}$ and remote
$z\paren{\v r, t}$ covariates impact a local spatio-temporal response
$Y\paren{\v s, t}$. The RESP model accounts for the influence of
covariates observed on a geographically remote domain $z\paren{\v r, t}$.

We demonstrate the capacity of the RESP model by
validating its ability to predict Colorado  winter precipitation in a
cross-validation study (\Cref{sec:caseintro}).  Our study represents a type of
perfect prognosis problem in which future precipitation will be studied
with covariates that have been simulated by GCMs.  Since
atmospheric processes have relatively short memory, it is reasonable to assume
winter precipitation is conditionally independent across years when local
and remote covariates are given.  Therefore, we develop the RESP model assuming
there is no meaningful temporal dependence.  We conclude with discussions of
temporal extensions and other directions for future work and further
application (\Cref{sec:discussion}).

\begin{knitrout}
\definecolor{shadecolor}{rgb}{0.969, 0.969, 0.969}\color{fgcolor}\begin{figure}

{\centering \includegraphics[width=\maxwidth]{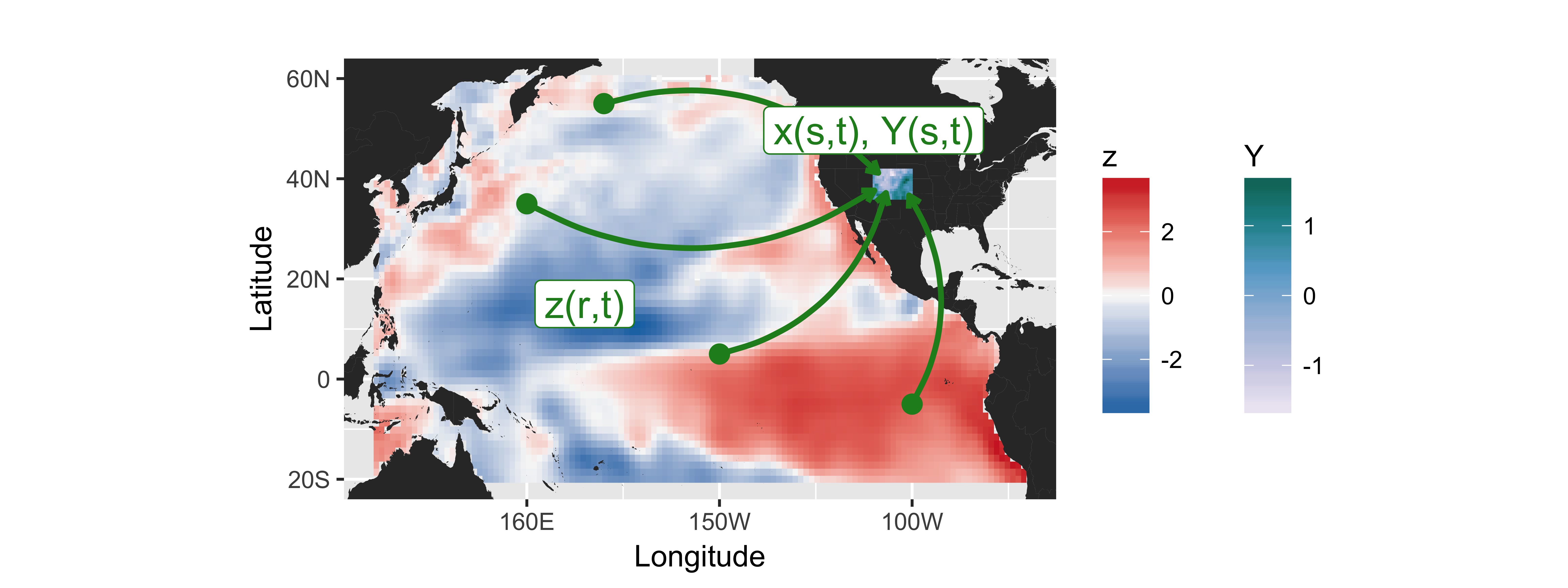} 

}

\caption{Schematic illustration of a teleconnection problem.  Colorado precipitation $Y\paren{\v s, t}$ is influenced by both local covariates $\v x\paren{\v s, t}$ and remote covariates $z\paren{\v r, t}$.  The remote covariates shown here are standardized anomalies of average monthly Pacific Ocean sea surface temperatures during Winter, 1982. The data come from the ERA-Interim reanalysis dataset \citep{Dee2011}.}\label{fig:tele_schematic}
\end{figure}

\end{knitrout}

\section{A geostatistical model for spatially remote covariates}
\label{sec:development}

Teleconnection manifests as an aggregate property of spatially continuous
covariates.  For example, consider the sea surface temperature (SST) at
location $\vr$ and time $t$, $z\paren{\vr, t}$.  In spatially varying
coefficient models \autoref{eq:soa_model}, it is common to adopt a
teleconnection index $z\paren t\in\mathbb R$ that is defined as the average SST
$z\paren{\vr, t}$ over a region $\mathcal R\subset\Dz$. In
\autoref{eq:soa_model}, the spatially varying coefficient
term $z\paren t \theta\paren\vs$ motivates the RESP model through the expansion
\begin{align}
\label{eq:ex_int}
  z\paren t \theta\paren\vs = \frac{1}{\card{\mathcal R}}
    \int_{\mathcal R} z\paren{\vr,t} \theta\paren\vs d\vr .
\end{align}

The RESP model extends the integral in \autoref{eq:ex_int} to the entire remote
domain $\Dz$ and allows $\theta\paren\vs$ to vary with respect to $\vr$,
distinguishing it from spatially varying coefficient models
(\Cref{sec:model}).  Integration is a natural construct for aggregating effects
of spatially continuous covariates, represents the conceptual limit of
studying teleconnection with increasingly fine subsets of $\mathcal R$,
and allows study of teleconnection with additional spatial structure and
without defining indices a priori.

\subsection{Model formulation}
\label{sec:model}

The remote effects spatial process (RESP) model extends the standard
geostatistical setting in which a local response variable
$Y(\v s, t) \in\mathbb R$ and known covariate vector
$\v x(\v s, t) \in \mathbb R^p$ are observable at discrete time points
$t \in \mathcal T = \brc{ t_1, \dots, t_\nt }$
and at locations $\vs$ in a continuous domain $\Dy$.  The RESP model includes
the effects of known remote covariates
$z(\v r, t) \in \mathbb R$, which are observable at locations $\vr$ in a
continuous domain that is spatially disjoint from the local response---i.e.,
in a continuous $\Dz$ s.t. $\Dy \cap \Dz = \emptyset$. The RESP model is given
by
  \begin{equation}
    Y(\v s, t) =
      \v x^T(\v s, t) \v\beta +
      w(\v s, t) +
      \varepsilon(\v s, t)  +
      \gamma\paren{\vs,t}
    \label{eq:stProcess}
  \end{equation}
where the regression coefficients $\v\beta \in \mathbb R^p$, spatially 
correlated noise $w(\v s, t)$, and independent noise $\varepsilon(\v s, t)$ are
standard components for spatial regression models
\citep[Chapters 6, 9, 11]{Banerjee2015}. In the RESP model the
teleconnection effect given by $\gamma\paren{\vs,t}$ is defined by
\begin{equation}
\label{eq:teleconnectionTerm}
  \gamma\paren{\vs,t} =
    \int_{\mathcal D_Z} z\paren{\vr,t} \alpha\paren{\vs,\vr} d\vr
\end{equation}
which describes the net effect of the remote covariates $z\paren{\vr,t}$ on
the continuous spatial process $Y\paren{\vs,t}$ at
discrete time $t$.  The integral \autoref{eq:teleconnectionTerm} reduces to a sum
for finite samples, in which the remote covariates $z\paren{\vr,t}$ are observed
at $\nr<\infty$ locations. Multivariate extensions of
\autoref{eq:teleconnectionTerm} are discussed in \Cref{sec:discussion}.

The remote (or teleconnection)
coefficients $\alpha\paren{\vs,\vr}$ are spatially correlated and doubly-indexed by
$\paren{\vs,\vr}\in\Dy\times\Dz$.  The spatial correlation and double-indexing
of $\alpha\paren{\vs,\vr}$ represents teleconnection effects that vary
regionally in the sense that the response $Y\paren{\vs,t}$ at one location
$\vs\in\Dy$ can respond to the remote covariates $z\paren{\vr,t}$ more
strongly than the response $Y\paren{\vs',t}$ at another location $\vs'\in\Dy$.
Similarly, the response $Y\paren{\vs,t}$ at one location $\vs\in\Dy$ can respond
differently to remote covariates $z\paren{\vr,t}$ and $z\paren{\vr',t}$ at distinct
remote locations $\vr,\vr'\in\Dz$.
Thus, the remote coefficients $\alpha\paren{\vs,\vr}$ vary spatially and use the
remote covariates $z\paren{\vr, t}$ to provide local adjustment to the mean
response.  The teleconnection term $\gamma\paren{\vs,t}$ is well
defined because we assume the remote covariates $z\paren{\vr,t}$ are known and
square-integrable over $\Dz$ at each time point $t$ \citep[Section 5.2]{Adler2007}.

The RESP model provides a simple geostatistical approach to modeling
teleconnections by extending spatial regression models to incorporate data from
spatially remote regions.
The teleconnection term $\gamma\paren{\vs,t}$ distinguishes the RESP model
\autoref{eq:stProcess} from standard geostatistical models, in which---for
example---the responses $Y\paren{\vs,t}$ and $Y\paren{\vs',t}$ at distinct
spatial locations $\vs,\vs'\in\Dy$ are only influenced by distinct covariates
$x\paren{\vs,t}$ and $x\paren{\vs',t}$.  To model the influence of
teleconnection phenomena the RESP model lets the remote covariates $z\paren{\vr,t}$
simultaneously influence the responses $Y\paren{\vs,t}$ and $Y\paren{\vs',t}$.

Geostatistical modeling conventions use mean zero Gaussian processes to specify
the randomness of the unknown, spatially correlated components $w(\v s, t)$ and
$\alpha(\v s, \v r)$, and an independent processes to specify the
noise $\varepsilon(\v s, t)$---the nugget.  We complete the Gaussian process
specifications by defining the
covariance functions for the spatially correlated components.  Let $C_w$ and
$C_\alpha$ respectively be the covariance functions for
$w(\v s, t) + \varepsilon(\v s, t)$ and $\alpha(\v s, \v r)$, where
\begin{align}
\label{eq:local_cov_def}
C_w \brc{\paren{\v s, t}, \paren{\v s', t'}} =&
  \paren{\kappa\paren{\v s, \v s'; \v\theta_w} +
    \sigma^2_\varepsilon \mathds{1}\paren{\v s = \v s'} }\mathds{1}\paren{t=t'},
\\
\label{eq:remote_cov_def}
C_\alpha \brc{\paren{\v s, \v r}, \paren{\v s', \v r'}} =&
  \paren{ \kappa\paren{\v s, \v s'; \v\theta_w} +
    \sigma^2_\varepsilon \mathds{1}\paren{\v s=\v s'} }
  \kappa\paren{\v r, \v r'; \v\theta_\alpha} .
\end{align}
Our model may be developed with any spatial covariance function $\kappa$, but
here we choose to work with the stationary Mat\'{e}rn covariance
\begin{align}
\kappa\paren{\vu,\vv; \v\theta} =
  \frac{\sigma^2}{2^{\nu-1}\Gamma\paren{\nu}}
  \paren{d\paren{\vu, \vv}/\rho}^\nu
  K_\nu\paren{d\paren{\vu, \vv}/\rho}
\label{eq:materndef}
\end{align}
for spatial locations $\vu$ and $\vv$, and parameter vector
$\v\theta = \paren{\sigma^2, ~\rho, ~\nu}^T$.  The function $d\paren{\vu,\vv}$
must be an appropriate distance function (e.g., great-circle distances for
locations on a sphere),
$\sigma^2>0$ is a scaling parameter, $\nu>0$ is a smoothness
parameter, $\rho>0$ is a range parameter, and $K_\nu$ is the modified Bessel
function of the second kind with order $\nu$.  In covariance function definitions
\autoref{eq:local_cov_def} and \autoref{eq:remote_cov_def}, $\mathds{1}$
represents the indicator function and $\sigma^2_\varepsilon$ represents
the variance of the nugget process which we specify to be a collection of
independent and identically distributed mean zero Gaussian random \\
variables---i.e.,
$\varepsilon(\v s, t) \overset{iid}{\thicksim} \mathcal N
  \paren{0, ~ \sigma^2_\varepsilon} ~
  \forall\paren{\vs,t}\in\Dy\times\mathcal T$.

While the definitions \autoref{eq:local_cov_def} and \autoref{eq:remote_cov_def}
for the local and remote covariances $C_w$ and $C_\alpha$ can be generalized,
the definitions restrict our use of the RESP model to working in
the perfect prognosis downscaling setting described at the end of
\Cref{sec:intro}.  The responses $Y\paren{\vs,t}$ and $Y\paren{\vs,t'}$ for
$t\neq t'$ are independent given covariates and
sufficiently separated time points, like successive winters
(e.g., winter 1991, winter 1992, etc.).  The remote covariates in the
teleconnection term \autoref{eq:teleconnectionTerm} naturally induce temporal
non-stationarity in the response's variance; extensions to accommodate serial
dependence are discussed in \Cref{sec:discussion}.
The remote covariance $C_\alpha$ also
induces a separable structure for the remote coefficients
$\alpha\paren{\vs,\vr}$, which constrains the spatial variability of
teleconnection effect fields and simultaneously constrains the
teleconnection effects $\brc{\alpha\paren{\vs,\vr}:\vr\in\Dz}$ and
$\brc{\alpha\paren{\vs',\vr}:\vr\in\Dz}$ to be similar for nearby locations
$\vs, ~\vs'\in\Dy$.  Simpler covariance structures for the
teleconnection effects $\alpha\paren{\vs,\vr}$ may not capture these
physical properties of teleconnection as directly.  Similarly, although climate
data are often available as gridded data products, we choose to work with
geostatistical covariance models
\citep[or their discrete approximations, e.g.,][]{Lindgren2011a} instead of
neighborhood-based spatial models so that we may avoid inducing potentially
counterintuitive covariance structures \citep{Wall2004, Assuncao2009}.

\subsection{Reduced rank approximation}
\label{sec:reducedrank}

To apply the RESP model \autoref{eq:stProcess}, additional constraints need to
be imposed due to the potential multicollinearity in the covariates. Remote
covariates $z\paren{\v r, t}$ in teleconnection applications will often consist
of data that measure ocean properties at high spatial resolution,
like sea surface temperature or sea level pressure.  This raises concerns for
estimating the remote coefficients $\alpha\paren{\vs,\vr}$ in
\autoref{eq:teleconnectionTerm} as the main trends in the remote covariates
$z\paren{\vr, t}$ are highly collinear over $\mathcal D_Z$.  Physically, however,
this suggests the remote coefficients should be highly correlated as well.
We use predictive processes to
mitigate multicollinearity in the remote covariates, which is an alternative
motivation for predictive processes. \citet{Banerjee2008a} originally introduce
predictive processes so that parameters of
geostatistical models can be estimated for large spatial datasets,
rather than as an approach for mitigating spatial multicollinearity.
We consider more general basis expansions of remote coefficients in
\Cref{sec:reparam}.

We assume the remote coefficients $\alpha\paren{\vs, \vr}$ can be well
represented by weighted averages of remote coefficients $\alpha\paren{\vs, \vr^*}$
at knot locations $\v r^*_1, \dots, \v r^*_k \in \mathcal D_Z$, so we
 make the simplifying approximation that, for some weight
function $h\paren{\v r, \v r'}$ and associated vector
$\v h^*\paren{\v r} = \brk{h\paren{\v r, \v r_j^*}}_{j=1}^k \in\mathbb R^k$,
we can write
  \begin{equation}
  \label{eq:alpha_approx}
    \alpha\paren{\v s, \v r} = \sum_{j=1}^k
      h\paren{\v r, \v r_j^*}
      \alpha\paren{\v s, \v r_j^*}
    =
      \v h^*\paren{\v r}^T \v\alpha^*\paren{\v s} ,
  \end{equation}
where $\v\alpha^*\paren{\v s} = \brk{\alpha\paren{\v s, \v r_j^*}}_{j=1}^k
\in\mathbb R^k$.
The predictive process approach uses kriging to motivate a choice for the weight
vector $\v h^*\paren{\v r}$, which induces a weight function $h$.
Using Gaussian processes  in \Cref{sec:model} to model
 the remote coefficients implies that $\alpha\paren{\v s, \v r}$ and
 $\v\alpha^*\paren{\v s}$ are jointly normally distributed, yielding the
conditional expectation for $\alpha\paren{\v s, \v r}$
  \begin{equation}
  \label{eq:pred}
    \E{\given{\alpha\paren{\v s, \v r}}{\v\alpha^*\paren{\v s}}} =
    \v c^* \paren{\v r}^T \inv{{R^*}} \v\alpha^*\paren{\v s}
  \end{equation}
in which $\v c^* \paren{\v r} = \brk{
  C_\alpha \brc{\paren{\v s, \v r}, \paren{\v s, \v r_j^*}}
  }_{j=1}^k \in\mathbb R^k$
and
  $R^* \in\mathbb R^{k\times k}$ with entries \\
  $R^*_{ij} = C_\alpha \brc{\paren{\v s, \v r_i^*}, \paren{\v s, \v r_j^*}}$.
Note that the assumption in \autoref{eq:remote_cov_def} that $C_\alpha$ is
stationary means that $\v c^* \paren{\v r}$ and
$R^*$ do not depend on $\v s$, despite the term appearing in their definitions.
The predictive process approach
uses the conditional expectation \autoref{eq:pred} to define the
weight vector $\v h^*\paren{\v r} = \inv{{R^*}} \v c^* \paren{\v r}$ in the
approximation \autoref{eq:alpha_approx}.  \citet{Banerjee2008a} show that these
types of approximations are reduced rank projections that can capture
large-scale spatial structures in data.

Beyond mitigating the statistical issue of multicollinearity, the predictive
process approach relates the RESP model to spatially varying coefficient models
\autoref{eq:soa_model} and also has a scientific interpretation for
teleconnection. Using the reduced rank approximation \autoref{eq:alpha_approx}
to manipulate the integral in \autoref{eq:stProcess} shows that the reduced
rank approximation can be interpreted as inducing transformed covariates
$z^*\paren{\vr^*, t}$ via
\begin{equation}
\begin{split}
  \int_\Dz z\paren{\vr,t} \alpha\paren{\vs,\vr} d\vr =&
  \int_\Dz
    z\paren{\vr,t}
    \sum_{j=1}^k h\paren{\v r, \v r_j^*} \alpha\paren{\v s, \v r_j^*}
    d\vr
  \\
 =&  \sum_{j=1}^k
  \alpha\paren{\v s, \v r_j^*}
  z^*\paren{\v r_j^*, t}
\end{split}
\end{equation}
where $z^*\paren{\v r_j^*, t} = \int_\Dz z\paren{\vr,t} h\paren{\v r, \v r_j^*}
d\vr$.  The $z^*\paren{\v r_j^*, t}$ and $\alpha\paren{\v s, \v r_j^*}$ may
be collected into the covariate vector $\v z\paren t$ and spatially varying
effects $\v\theta\paren\vs$ in \autoref{eq:soa_model}.  We remark that the RESP
model differs from standard
spatially varying coefficient models in that the $z^*\paren{\v r_j^*, t}$
represent induced---rather than a priori---covariates, and the
$\alpha\paren{\v s, \v r_j^*}$ inherit spatial structure from the model's
formulation.

Scientifically, the predictive process approach to addressing multicollinearity
in the remote covariates reduces the remote covariates $z\paren{\vr, t}$,
$\vr\in\Dz$ at each time point to $k$ spatially-averaged indices
$z^*\paren{\vr^*, t}$ centered at $\vr^*$ for
$\vr^*\in\brc{\vr_1^*, \dots, \vr_k^*}$.
This manipulation is fairly generic and should be applicable
to all predictive process models.  For teleconnection, this manipulation
connects the RESP model to one set of standard teleconnection methodologies in
which teleconnection effects are measured with respect to ocean indices based
on spatial averages of remote covariates \citep{Ashok2007, Towler2016}.

\subsection{Spatial basis function transformation of remote coefficients}
\label{sec:reparam}

The RESP model \autoref{eq:stProcess} is also related to another set of
standard teleconnection methodologies in which teleconnection effects are
measured with respect to complex ocean indices such as empirical orthogonal
functions \citep{Ting1997, Montroy1997}.
Spatial basis functions provide a means to reparameterize the RESP model and
show it can identify and leverage known teleconnections with
complex patterns.  We use the following reparameterization of
the teleconnection effects $\alpha\paren{\vs,\vr}$ to discuss teleconnection
between Pacific Ocean sea surface temperature and Colorado precipitation in
\Cref{sec:caseintro}.

Complex teleconnection patterns are often based on spatial basis function
expansions of the remote covariates $z\paren{\vr, t}$. If there exist weights
$\brc{a_l\paren{t}: l=1,\dots,K;~ t\in\mathcal T}$ such that the remote covariates
$z\paren{\vr, t}$ can be written as a linear combination of continuous,
time-invariant basis functions
$\brc{\psi_l\paren{\vr}:l=1,\dots,K;~ \vr\in\Dz}$ via
\begin{equation}
\label{eq:remote_expansion}
  z\paren{\vr, t} = \sum_{l=1}^K a_l\paren{t} \psi_l\paren{\vr} ,
\end{equation}
then linearity of the integral in \autoref{eq:teleconnectionTerm} and reduced rank
approximation \autoref{eq:alpha_approx} can induce a reparameterized, reduced-rank
teleconnection effect process $\alpha'\paren{\vs, l}$ for patterns
$l=1,\dots, K$ by
\begin{equation}
\label{eq:reparam}
  \alpha'\paren{\vs, l} =
    \sum_{j=1}^k \alpha\paren{\vs, \vr_j^*}
                 \int_\Dz \psi_l\paren{\vr} h\paren{\vr, \vr_j^*} d\vr ~.
\end{equation}
Note that the transformation appears naturally because
\begin{equation}
\label{eq:reparam_motivation}
\begin{split}
  \int_\Dz z\paren{\vr,t} \alpha\paren{\vs,\vr} d\vr =&
  \int_{\Dz}
    {\sum_{l=1}^K
      a_l\paren{t}
      \psi_l\paren{\vr}
    }
    {\sum_{j=1}^k
      h\paren{\vr,\vr_j^*}
      \alpha\paren{\vs,\vr_j^*}
    }
    d\vr
  \\ =&
  \sum_{l=1}^K
    a_l\paren{t}
    \sum_{j=1}^k
      \alpha\paren{\vs,\vr_j^*}
      \int_{\Dz}
        \psi_l\paren{\vr}
        h\paren{\vr,\vr_j^*}
        d\vr
  \\ =&
  \sum_{l=1}^K a_l\paren{t} \alpha'\paren{\vs, l} .
\end{split}
\end{equation}
As with the reduced rank approximation \autoref{eq:alpha_approx}, the
transformation \autoref{eq:reparam_motivation} also relates the RESP model
to spatially varying coefficient models \autoref{eq:soa_model} and has
scientific relevance for teleconnection.  The deterministic remote covariate
weights $a_l\paren{t}$ and reparameterized remote coefficients
$\alpha'\paren{\vs, l}$ may be collected into the covariate vector
$\v z\paren t$ and spatially varying effects $\v\theta\paren\vs$ in
\autoref{eq:soa_model}.  While the covariate weights $a_l\paren{t}$ suggest
a priori selection of teleconnection indices, the reparameterization may be
applied after model estimation. The $\alpha'\paren{\vs, l}$ additionally
inherit spatial structure from the model's formulation.    Scientifically,
a special case of \autoref{eq:remote_expansion} are principal component
decompositions or the closely related truncated Karhunen--L\`{o}eve expansions,
which are referred to as empirical orthogonal functions (EOFs)
in climate science.  EOFs are particularly useful expansions for teleconnection
because these transformations meaningfully characterize phenomena that impact
global climate \citep{Ashok2007}.

\subsection{Inference}
\label{sec:resid_interp}

While inference for the RESP model \autoref{eq:stProcess} can use standard
hierarchical Bayesian modeling techniques, the Bayesian framework provides
crucial intuition and interpretation for estimates of teleconnection effects
\autoref{eq:alpha_approx} and \autoref{eq:reparam}.   Full description of model
priors and computational techniques for inference are discussed in Supplement A
\nolink{\Cref{supp:sec:detailed_implementation}}.
The Gaussian process assumption and separable covariance
\autoref{eq:remote_cov_def} for the vector of teleconnection coefficients
$\v\alpha^*(\vs)$ with associated covariance matrix $R^*$ defined in
\Cref{sec:reducedrank} imply the normally-distributed prior
$\given{\v\alpha^*\paren{\v s}}{R^*} \thicksim \mathcal N\paren{\v 0, R^*}$.
Gaussian process assumptions for the RESP model's spatial correlation also
imply the likelihood for the vector of responses observed at $\nt$ timepoints
$\v Y\paren\vs = \brk{Y\paren{\vs, t_1}, \dots, Y\paren{\vs, t_\nt}}^T
  \in\mathbb R^\nt$ is
 \begin{align}
    \given{\v Y\paren{\v s}}
      {\v\alpha^*\paren{\v s}, \v\beta, R^*, \v c^*, \sigma^2_\vs}
      \thicksim \mathcal N \paren{
        \v X\paren{\v s} \v \beta + {\v Z^*}^T \v\alpha^*\paren{\v s},~
        \sigma^2_\vs I_\nt
      }
\end{align}
with $\sigma^2_\vs = C_w \brc{\paren{\vs, t}, \paren{\vs, t}}$ and matrices of
local covariates
 $\v X\paren{\v s} = \brk{\v x\paren{\v s, t}^T}_{t=t_1}^{t_\nt }
    \in\mathbb R^{\nt \times p}$
and reduced-rank remote covariates $\v Z^* \in\mathbb R^{k\times\nt}$. The
matrix $\v Z^*$ is comprised of column vectors
$\v z_t^* = \inv{{R^*}} {\v{c}^*}^T \v z_t \in\mathbb R^k$ built from
remote covariate vectors
$\v z_t = \brk{ z\paren{\v r_j, t}}_{j=1}^\nr \in \mathbb R^\nr$.
Our formulation of the spatial correlation
\autoref{eq:local_cov_def} implies the scalar $\sigma^2_\vs$ is constant across
time; non-stationary extensions are discussed in \Cref{sec:discussion}.
Standard Bayesian linear regression results \citep[Example 5.2]{Banerjee2015}
yield the posterior distribution
\begin{align}
  \given{\v\alpha^*\paren{\v s}}
    {\v Y\paren{\vs}, \v\beta, R^*, \v c^*, \sigma^2_\vs}
  \thicksim \mathcal N \paren{
    \sigma^{-2}_\vs \v\Psi \v Z^*
    \paren{\v Y\paren\vs - \v X\paren{\v s} \v \beta}
    ,~ \v\Psi
  }
\end{align}
for
\begin{align*}
  \v\Psi = \inv{\paren{ {\inv{ {R^{*}}}} + \sigma^{-2}_\vs \v Z^* {\v Z^*}^T }} .
\end{align*}

The connection to Bayesian linear regression lends intuition for inference
on the remote effects $\v\alpha^*\paren{\v s}$.  In particular, the connection
provides intuition for using the RESP model when some local covariates
$\v x\paren{\vs ,t}$ are also teleconnected with remote covariates $\v z_t$. Remote
coefficients can be interpreted as residual teleconnection effects in the sense
that they model the impact of remote covariates on the response after removing
local effects
$\v X\paren{\v s} \v \beta$. Properties of regressions also imply patterns in
maps of posterior means for $\v\alpha^*\paren{\v s}$ may resemble patterns in
maps that show pointwise correlations $\Cor_t\paren{z^*(\v r^*, t), Y(\v s, t)}$
between remote covariates at $\vr^*$ and responses at $\vs$.  Similar
regression-based interpretations can be derived for the reparameterized
teleconnection coefficients \autoref{eq:reparam}.

\section{Climate application: Colorado winter precipitation}
\label{sec:caseintro}

The RESP model \autoref{eq:stProcess} is applied here using remote and 
local covariates to estimate Colorado winter precipitation. Winter 
precipitation is important to estimate because it strongly in influences 
Colorado's annual water supply. We investigate the utility of our RESP model 
for this application because there is considerable uncertainty regarding 
precipitation that is directly predicted by GCMs. Further, the RESP model can 
be applied without specifying teleconnection indices a priori, as many common 
approaches require. Let $Y\paren{\vs, t}$ denote
average monthly precipitation in winter for location $\vs$ and year $t$ via
\begin{align}
  Y\paren{\vs, t} = \paren{
    Y_{Dec}\paren{\vs, t} +
    Y_{Jan}\paren{\vs, t} +
    Y_{Feb}\paren{\vs, t} } / 3
\end{align}
in which, for example, $Y_{Dec}\paren{\vs, t}$ represents the total
December precipitation in year $t$ at location $\vs$. The atmosphere's short
memory suggests $Y\paren{\vs,t}$ is independent from $Y\paren{\vs,t'}$ for
$t\neq t'$, which is confirmed in an exploratory analysis of Colorado
precipitation. Winter precipitation
is important to estimate at long time scales because it strongly influences
Colorado's annual water supply.

We formulate the problem of estimating precipitation
as a need to estimate entire precipitation fields when only covariates
are available. We build the
RESP model \autoref{eq:stProcess} with historical data to estimate a statistical
relationship between average monthly winter precipitation in Colorado and land
and sea surface temperatures.  
We discuss inference for the RESP model to illustrate that it can
estimate teleconnection patterns without specifying teleconnection indices a
priori (\Cref{sec:inference}). A leave-one-out cross-validation study validates
the model's effectiveness (\Cref{sec:predresults}), especially in relation to
other common downscaling methods (\Cref{sec:comparison_description}).
Although beyond the scope of this study, a next step for future work would be 
to apply the RESP model to simulated GCM output.

\subsection{Data}
\label{sec:data}

The ERA-Interim reanalysis dataset provides reconstructions of historical
sea surface temperatures and local covariates \citep{Dee2011}.  The response,
precipitation, comes from the PRISM dataset \citep{Daly2008}.
We limit our study period to 1981 through 2013 because earlier
records of large scale climate are less complete. Both datasets are
reanalysis products, which are necessary because working directly with
observations can be challenging. Raw data may be from various sources and are
often spatially sparse and temporally incomplete.
Reanalysis products use statistical techniques and physical relationships to
reproduce consistent datasets at regular, gridded locations with complete
records after removing or correcting observations that are physically
inconsistent or from stations with potential data collection issues.

This study uses data averaged over the boreal winter months (December,
January, February) because Northern Hemisphere teleconnections are often
strongest in winter \citep{Nigam2015}.  We simplify the demonstration
using spatially-referenced variables average surface air temperature over Colorado
($T$) and average Pacific Ocean sea surface temperatures ($SST$) between
$120^\circ$E--$70^\circ$W and $20^\circ$S--$60^\circ$N to predict the
spatially-referenced
response, average winter precipitation in Colorado ($P$). We standardize
all data to remove the impact of orographic and other location-based effects
by removing the pointwise mean from all data and scaling data to have unit
variance.  We additionally scale the $SST$ values by $\nr^{-1}$ to ensure the
remote coefficient magnitudes are independent of the resolution at which $SST$
is measured.  We standardize our data before conducting the leave-one-out
cross-validation study so all of the testing and training data are
comparable. Thus, our data are standardized climate anomalies that, for
example, represent the number of standard deviations $P\paren{\vs,t}$ is above
or below the time-averaged value  $E_t\left[P\paren{\vs,t}\right]$ at
location $\vs$.
The data are also spatially aggregated so that $\ns=240$, 42 km-resolution grid
cells cover Colorado and $\nr=5,252$, 78 km-resolution grid cells cover the
Pacific Ocean.  Distances
between grid cells are measured with great-circle distances.
We spatially aggregate the PRISM data to increase the smoothness of the data
and to make the problem computationally tractable. We discuss alternate
approaches to improve computational tractability in \Cref{sec:discussion}.
The spatial aggregation and standardization also increase the normality of the
data and provide a scale for precipitation with negative support, making it
more appropriate for analysis with the RESP model's Gaussian likelihood.

\begin{knitrout}
\definecolor{shadecolor}{rgb}{0.969, 0.969, 0.969}\color{fgcolor}\begin{figure}

{\centering \includegraphics[width=\maxwidth]{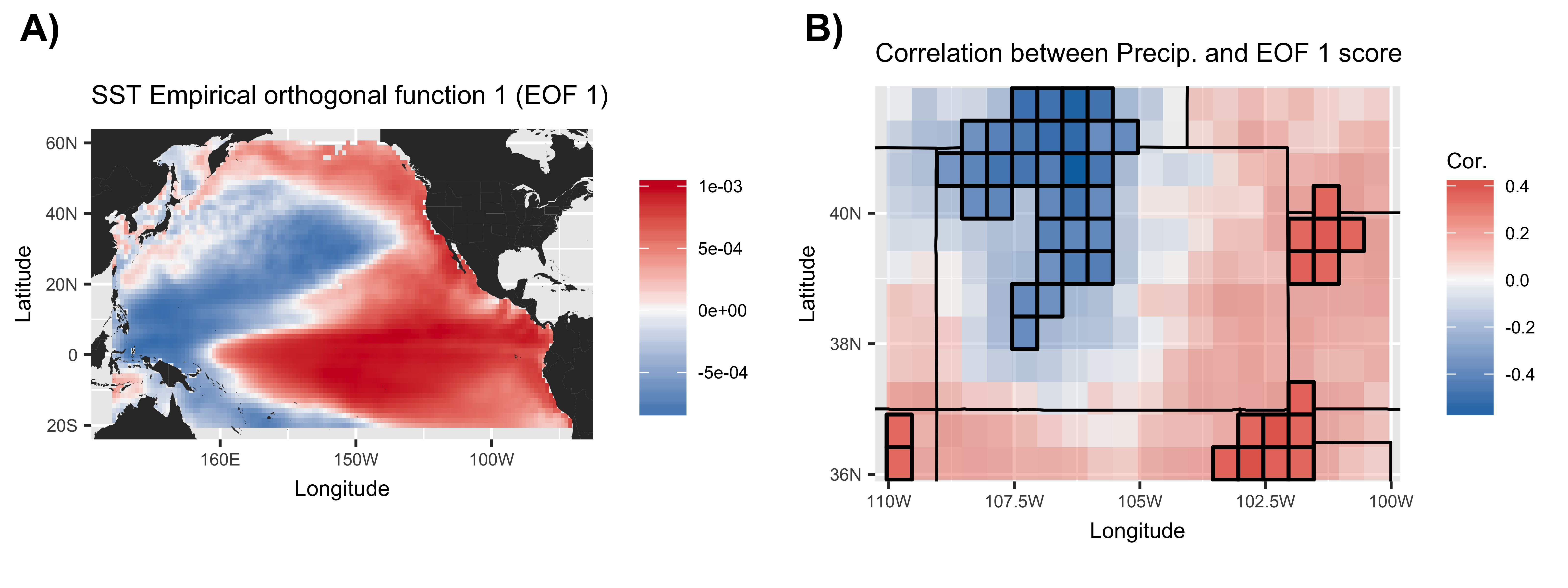} 

}

\caption{Exploratory analysis plots.  A) The first empirical orthogonal function (EOF) $\psi_1:\Dy\rightarrow\mathbb R$ for standardized anomalies of Pacific Ocean sea surface temperatures is an indicator of El Ni\~{n}o events, during which sea surface temperatures are anomalously warm in the central and eastern Pacific Ocean tropics but anomalously cool in the western tropics \citep{Ashok2007}.  EOF 1 accounts for 30\% of the variability in sea surface temperatures.  B) Pointwise correlations $\Cor_t\paren{P\paren{\vs,t}, a_1\paren{t}}$ between Colorado precipitation $P\paren{\vs,t}$ and the EOF 1 score $a_1\paren{t}$ suggest northern and western/central Colorado tends to receive less precipitation than average during El Ni\~{n}o events while eastern Colorado tends to receive more precipitation.  Significant correlations ($\text{naive independent p-value} <.05$) are highlighted, while non-significant correlations are faded slightly.}\label{fig:tele_eda}
\end{figure}

\end{knitrout}

Pacific Ocean sea surface temperature capture how the ocean influences
Colorado precipitation through the El Ni\~{n}o--Southern Oscillation (ENSO)
teleconnection \citep[Figure 2.4]{Lukas2014}.
The ENSO teleconnection is characterized by sea
surface temperatures that are anomalously warm in the central and eastern Pacific
Ocean tropics but anomalously cool in the western tropics. The first empirical
orthogonal function (EOF; i.e., principal component)
$\psi_1:\Dy\rightarrow\mathbb R$ for
Pacific Ocean sea surface temperature anomalies illustrates this pattern
(\autoref{fig:tele_eda}). Pointwise correlations
$\Cor_t\paren{a_1(t), P(\vs,t)}$ between the ENSO teleconnection's strength
$a_1\paren{t}$ and Colorado precipitation $P\paren{\vs, t}$ provide
standard evidence for teleconnection, suggesting northern and
western/central Colorado tend to receive significantly less precipitation than
average during ENSO events, which are periods of strong El Ni\~{n}o activity,
while plains regions bordering eastern Colorado tend to receive significantly
more precipitation than average (\autoref{fig:tele_eda}).

\subsection{RESP model and prior specification}
\label{sec:priors}

In the RESP model \autoref{eq:stProcess}, we specify a linear
relationship between the local covariate $T$ and response $P$ so that
$\v\beta$ in \autoref{eq:stProcess} has intercept $\beta_0$ and slope $\beta_T$
components $\v\beta = \paren{\beta_0, ~\beta_T}^T$.  While the RESP model as
described in \Cref{sec:model} uses a stationary covariance model and
precipitation is non-stationary in space, stationary models have comparable
predictive performance in Colorado \citep{Paciorek2006}.
For the RESP model's remote coefficients, knots are placed at 93 locations that
are roughly evenly spaced across the Pacific Ocean and along coastal locations
(Supplement A, \nolink{\autoref{supp:fig:knot_locations}}).
While knot selection can be problematic,
\citet{Banerjee2008a} find that reasonably dense, regularly spaced grids can
yield good results.  Since the ENSO teleconnection is scientifically meaningful,
we will interpret the transformed teleconnection effects
$\alpha'\paren{\vs,1}$ from \autoref{eq:reparam}, which are associated with
ENSO through its connection to the first empirical orthogonal function (EOF) of
sea surface temperature anomalies $\psi_1:\Dy\rightarrow\mathbb R$.

We adopt a combination of weakly informative and non-informative prior
distributions.  A dispersed normal prior is used for the fixed effects
 $\v\beta \thicksim \mathcal N\paren{\v 0, 10 I}$.
We use $\sigma_w^2\thicksim IG \paren{2,1}$,
$\sigma_\varepsilon^2\thicksim IG \paren{2,1}$,
$\rho_w \thicksim U\paren{1, 600}$, and $\rho_\alpha \thicksim U\paren{1, 2000}$.
The Mat\'{e}rn covariance smoothness parameters \autoref{eq:materndef} are fixed
at $\nu_w = \nu_\alpha = .5$, which correspond to the smoothest well-defined
Mat\'{e}rn covariances for Gaussian processes on spheres \citep{Gneiting2013}.
In exploratory analysis, variograms for the local and remote data fit this
parameterization well.
The prior for $\sigma^2_\alpha$ is informative to increase the identifiability
of this parameter and the remote range $\rho_\alpha$ \citep{Zhang2004a}.  The
prior $\sigma_\alpha^2\thicksim IG \paren{6,10}$ keeps the model from exploring
parameter combinations that would imply very large teleconnection influence
relative to the scale of the data $Y\paren{\vs, t}$.

\subsection{Comparison models}
\label{sec:comparison_description}

We demonstrate the benefit of remote covariates by comparing the RESP model
to RE and SP submodels that, respectively, exclude local and remote covariates.
We also show improvement to statistical downscaling and
prediction by comparing RESP model validation scores to spatially varying
coefficient (SVC) models \autoref{eq:soa_model} and other common downscaling
models, including a hybrid local and non-local regression using the
El-Ni\~{n}o--Southern Oscillation teleconnection (ENSO-T)
\citep[Sections 8.4, 8.5]{VandenDool2007}, canonical correlation analysis (CCA)
\citep[Chapter 14]{VonStorch1999}, and a baseline climatological
reference prediction (CLIM) \citep[Section 8.1]{VandenDool2007}.

While analog models provide an alternate means to model teleconnected processes,
we do not make comparisons to them in this application because analog models
require more temporal replication than our data provide.
Analog models require considerable temporal replication because predictions are
weighted combinations of past observations, where the weights are based on
distances between covariates at the prediction timepoint and all past
observations \citep{McDermott2016}.  An advantage of analog forecasts, for
example, is that the reweighting scheme naturally generates forecasts that
have the same spatial patterns as past observations.  Without enough past
observations, however, the likelihood increases that past observations are not
diverse enough to sufficiently approximate future states \citep{VanDenDool1994}.

\subsubsection{Spatially varying coefficient model (SVC)}

To facilitate comparison, the SVC model \autoref{eq:soa_model} is specified
with the same linear relationship between the local
covariate $T$ and response $P$ we use with the RESP model.  The scores
$a_1\paren t$ and $a_2\paren t$ for the first and second sea surface 
temperature (SST) anomaly EOFs $\psi_1$, $\psi_2$ capture ENSO and ENSO-Modoki 
teleconnection relationships for Colorado precipitation with bivariate 
spatially varying coefficients $\v\theta\paren\vs\in\mathbb R^2$. The scores
$\brc{a_i\paren t : i=1,2, ~ t\in\mathcal T}$ quantify the strength of ENSO
activity and are similar to other measures of ENSO activity \citep{Ashok2007}.
The first and second EOFs $\psi_1$ and $\psi_2$ respectively account for 30\% 
and 15\% of the variability in SST.

We adopt a hierarchical Bayesian framework to estimate the SVC model
\citep[Section 9.6.2]{Banerjee2015}.  An Inverse-Wishart
prior $\Lambda \thicksim IW\paren{I, 2}$ is used for
$\Lambda = \textrm{Cov}\paren{\v\theta\paren\vs}$ and a dispersed normal prior
is used for the fixed effects $\v\beta\thicksim\mathcal N\paren{\v 0, 10I}$.
We use $\sigma^2 \thicksim IG\paren{2,1}$ and $\rho \thicksim U\paren{1,600}$
for the prior distribution of the Mat\'{e}rn covariance with fixed smoothness
$\nu=.5$ for the model's spatial correlation.

\subsubsection{Hybrid local and non-local regression (ENSO-T)}

Pointwise regression models are commonly used to downscale climate data
\citep[e.g.,][]{Towler2016}.  The
ENSO-T model predicts precipitation $P\paren{\vs, t_0}$ at a location $\vs$ and
new time point $t_0$ by applying a regression of training data $P\paren{\vs, t}$
onto local surface air temperature $T\paren{\vs,t}$ and the score
$a_1\paren{t}$ for the first sea surface temperature EOF
$\psi_1:\Dy\rightarrow\mathbb R$. The ENSO-T downscaler provides a comparison
model that accounts for both local and remote effects, but not spatial dependence.

\subsubsection{Canonical correlation analysis (CCA)}

Canonical correlation analysis uses the empirical correlation structure of
sea surface temperature $SST$ and precipitation $P$ vectors to linearly map
these variables to a space in which the transformed vectors are maximally
correlated \citep[Chapter 14]{VonStorch1999}.  This mapping may be used in a
multivariate regression context with sea surface temperatures at new time points
to predict precipitation.  The mapping is often developed with some amount of
smoothing by removing higher order EOFs from the data.  We retain 16 EOFs in
our use of CCA because this lets us capture approximately 90\% of the variability
in the predictors $SST$ and predictand $P$.  The CCA downscaler provides a
comparison model that only accounts for remote effects and indirectly accounts
for spatial dependence.

\subsubsection{Climatological reference (CLIM)}

Climatologists use the unconditional distribution of
precipitation $P\paren{\vs,t}$ at a location $\vs$.  When no other
information is available, the average value of precipitation
$\textrm{E}_t\brk{P\paren{\vs,t}}$ is used as a climatological
point prediction for precipitation, and the empirical distribution is used
for probabilistic predictions. The CLIM downscaler provides a baseline
comparison model that does not account for spatial dependence, local, or remote
effects.

\subsection{Results}
\label{sec:results}

Model results are based on 20,000 samples from the posterior distribution after
a burn in period of 1,000 samples.  Convergence was assessed by examining
trace plots, autocorrelation plots, and effective sample sizes in addition to
comparing results from multiple runs with randomly initialized parameters.
Model adequacy was assessed using residual and qq-normal plots. These
diagnostics suggest there are no serious violations of the convergence and
distributional assumptions. Variance inflation factors (VIFs) that account for
the RESP model design also show no concern for multicollinearity in the fitted
model \nolink{\Cref{supp:sec:assessment}}.

\subsubsection{Inference}
\label{sec:inference}

Parameter estimates for the RESP model yield reasonable scientific
interpretations (\autoref{table:param_ests}).  The sign of the regression
coefficient $\beta_T$ for the temperature covariate $T$ is consistent with
physical processes that influence precipitation \citep{Daly2008}.  The local
covariance range parameter $\rho_w$ implies the dependence between locations
$\v s\in\mathcal D_Y$ has an effective range between 500 and 570 km, which is
the distance between locations beyond which the Mat\'{e}rn correlation
\autoref{eq:materndef} is small ($\leq .05$).  This length scale is in the
size range of mesoscale weather processes that produce precipitation
\citep{Parker2015a}. The remote covariance range parameter $\rho_\alpha$ implies
the dependence between locations $\v r\in\Dz$ has an effective range between
720 and 2,200 km, which is roughly the size of the mid-sized structures seen
in the EOF patterns in \autoref{fig:tele_eda} A.   Since local temperature $T$
is teleconnected with sea surface temperatures $SST$, remote effects
must be interpreted as residual teleconnection effects, as described at the end
of \Cref{sec:resid_interp}.  Significant remote effects suggest Colorado's
teleconnection with the Pacific Ocean cannot be represented through a linear
relationship with temperature alone;  the teleconnection likely involves
non-linear relationships and additional variables or interactions.
Posterior estimates for the transformed remote
effects $\brc{\alpha'\paren{\vs, 1}:\vs\in\Dy}$ associated with
$\psi_1:\Dy\rightarrow\mathbb R$ (\autoref{fig:teleconnection_estimates})
largely match the exploratory pointwise correlations between $P\paren{\vs,t}$
and $a_1\paren{t}$ found in the exploratory plot (\autoref{fig:tele_eda}),
indicating the RESP model \autoref{eq:stProcess} is capturing known Colorado
teleconnections.  Fewer locations have significant teleconnection, however, as
the estimates incorporate more uncertainty due to spatial correlation;
significance is determined with respect to evaluating highest posterior
density intervals, separately for each location $\vs\in\Dy$.

\begin{table}[t!]
\centering
\caption{Posterior mean estimates and 95\% highest posterior density (HPD) intervals for the RESP model's parameters, which include an intercept $\beta_0$ and temperature effect $\beta_T$ on the mean response (see equation \autoref{eq:stProcess}), and covariance scale $\sigma^2$ and range $\rho$ parameters for the local $w$ and remote $\alpha$ spatial dependence and nugget effect $\varepsilon$  (see \autoref{eq:local_cov_def} and \autoref{eq:remote_cov_def}).  The smoothness parameters $\nu_w$ and $\nu_\alpha$ were fixed (\Cref{sec:priors}).} 
\label{table:param_ests}
\begin{tabular}{rlrc}
  \toprule
 &  & Posterior mean & 95\% HPD \\ 
  \midrule
Local effects & $\beta_0$ & $-$0.00 & ($-$0.14, 0.14) \\ 
    & $\beta_T$ & $-$0.18 & ($-$0.24, $-$0.12) \\ 
   \midrule
   & $\sigma^2_w$ & 0.55 & (0.49, 0.62) \\ 
      & $\sigma^2_\alpha$ & 6.05 & (1.04, 14.81) \\ 
  Covariance & $\sigma^2_\varepsilon$ & 0.01 & (0.01, 0.01) \\ 
       & $\rho_w$ & 248.00 & (220, 280) \\ 
        & $\rho_\alpha$ & 509.00 & (266, 799) \\ 
   \bottomrule
\end{tabular}
\end{table}

%\FloatBarrier

\begin{knitrout}
\definecolor{shadecolor}{rgb}{0.969, 0.969, 0.969}\color{fgcolor}\begin{figure}

{\centering \includegraphics[width=\maxwidth]{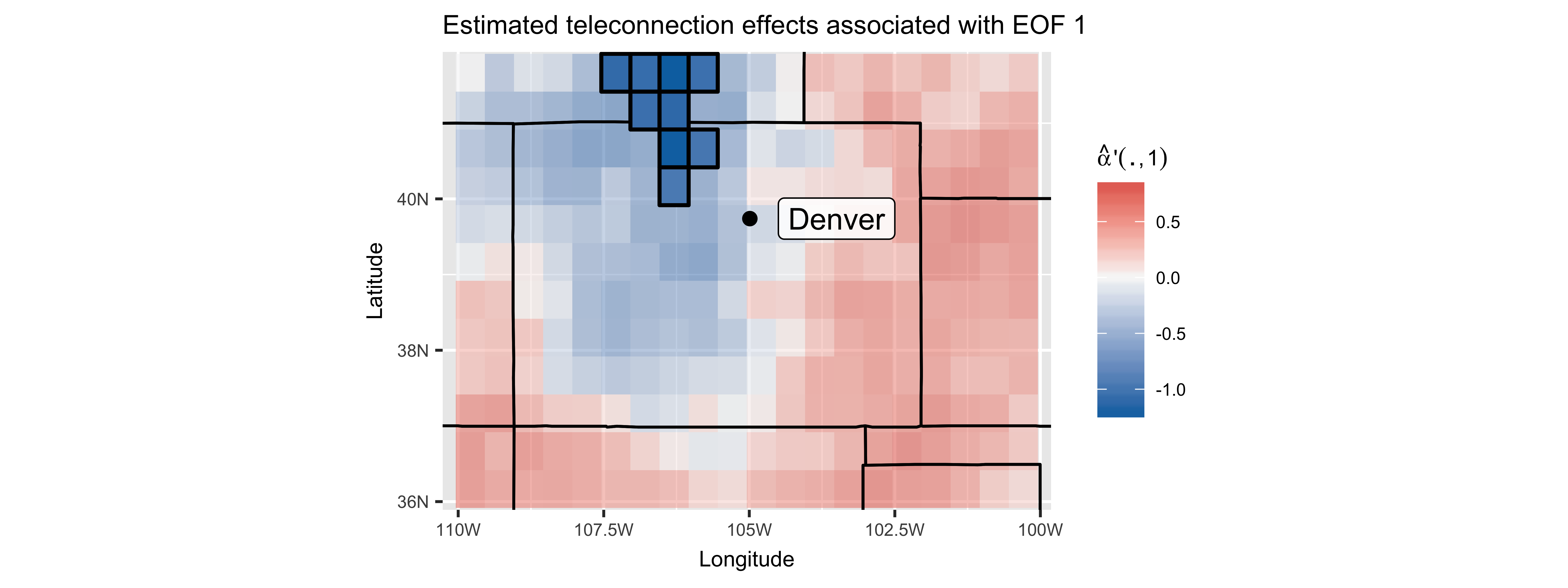} 

}

\caption{Estimated teleconnection effects $\hat\alpha'\paren{\vs, 1}$ for EOF 1 $\psi_1:\Dy\rightarrow\mathbb R$.  The overall patterns yield similar interpretations as those made with the \autoref{fig:tele_eda} exploratory plots, however, the RESP model reduces the regions in which evidence exists for significant teleconnection. Significant teleconnection effects, as determined using 95\% highest posterior density intervals, are highlighted.}\label{fig:teleconnection_estimates}
\end{figure}

\end{knitrout}

\subsubsection{Model validation}
\label{sec:predresults}

Leave-one-out cross-validation scores demonstrate the RESP model benefits from
including remote covariates and offers improvement over comparison models in
the intended prediction-like setting of perfect prognosis downscaling
(\autoref{fig:crps_skill}). The RESP
and comparison models are trained on all but one year of available data, then
used to predict the responses $\brc{P\paren{\vs, t} : \vs\in\Dy }$ for the test
year $t$ to mimic the perfect prognosis downscaling setting in which a climate
variable must be completely inferred from covariate data only.  The process is
repeated with all years of available data.  While the RESP
and comparison models yield continuous predictive distributions, we
discretize the distributions before assessing them.  Climate forecasts are
often discretized because it is inherently difficult to develop more precise
climate predictions at seasonal and longer time scales
\citep[][Section 9.6]{Mason2012, VandenDool2007}.  We use the empirical
terciles $\hat q\paren{1/3; \, P\paren{\vs, \cdot} }$ and
$\hat q\paren{2/3; \, P\paren{\vs, \cdot} }$ to discretize the predictive
distribution $f\paren{ \left. P\paren{\vs, t_0} \right\vert \v P}$ at each
location $\vs\in\Dy$ into ``below average'', ``near average'', and
``above average'' categories.  While it is possible to directly fit discrete
models to the data, doing so is not necessarily helpful.  For example, a
probit-link RESP or SVC model would require re-estimation of observed
continuous data $P\paren{\vs, t}$ as latent fields \citep{Higgs2010}.

We use ranked probability scores (RPS) to assess probabilistic forecasts for
ordinal variables, giving lower scores to models that generate predictive
distributions that better match the true distribution \citep{Gneiting2007a}.
The CCA model only yields point
predictions since predictive uncertainties are difficult to obtain.  Thus, the
CCA's validation scores are inflated since its discretized predictive
distribution is defined by a point mass on the category that matches the
tercile in which the point prediction lies.

The RESP model \autoref{eq:stProcess} frequently yields better
probabilistic predictions than the comparison models.  In particular, the RESP
model performs better than the RE or SP submodels which highlights the
advantage of combining local and remote information.  Sample maps of predictions
and uncertainties are presented in Supplement A \nolink{\Cref{supp:predmapapp}}.
The RESP model also tends
to perform better than the SVC model which highlights the advantage of not
specifying teleconnection indices a priori and adding additional spatial
structure to estimates of teleconnection effects.
Similar results are obtained using Heidke skill scores to compare models.
Heidke skill scores are commonly used in climate science to
measure a model's misclassification rate for categorical point predictions
\citep[Section 18.1]{VonStorch1999}.  Formulas and details for RPS and Heidke
skill scores can be found in Supplement A \nolink{\Cref{supp:sec:assessment}}.

\begin{knitrout}
\definecolor{shadecolor}{rgb}{0.969, 0.969, 0.969}\color{fgcolor}\begin{figure}

{\centering \includegraphics[width=\maxwidth]{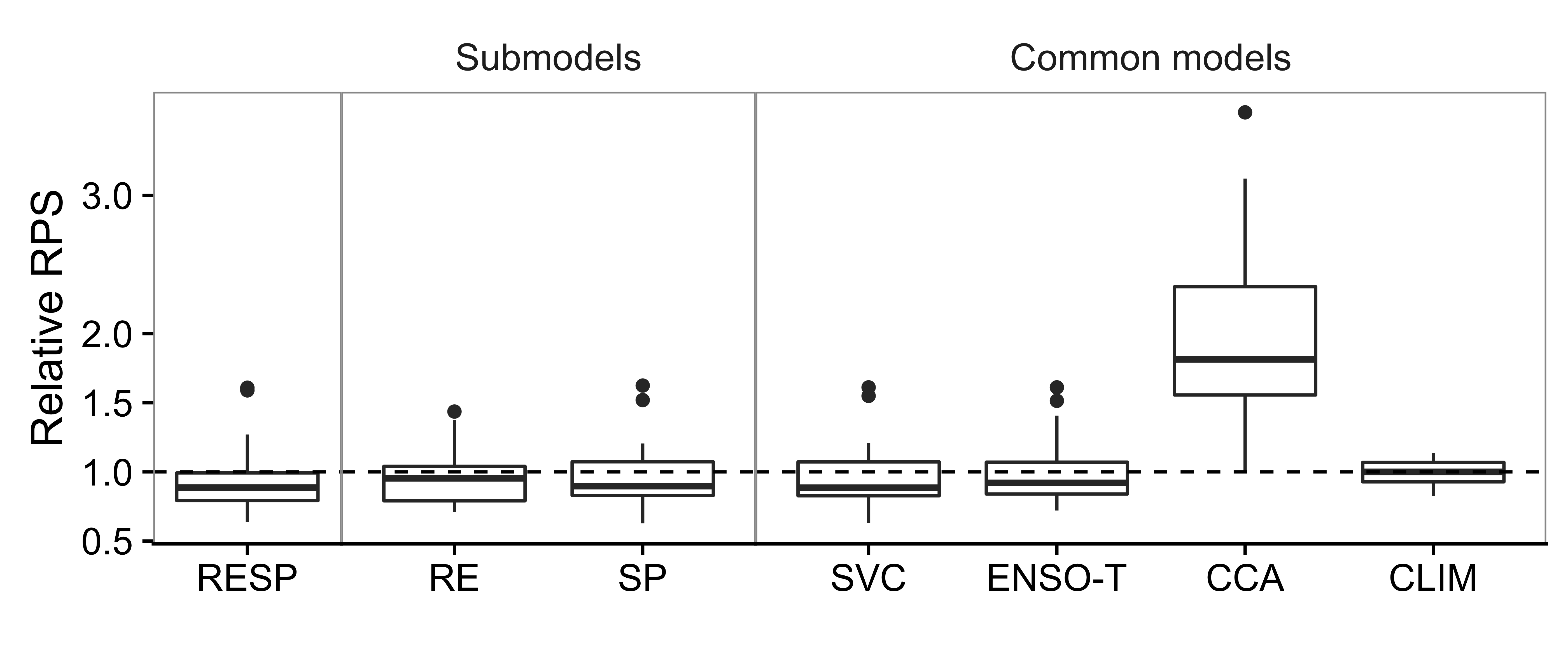} 

}

\caption[Comparison of Ranked probability scores (RPS) for probabilistic categorical predictions on the leave-one-out test datasets for the RESP and comparison models]{Comparison of Ranked probability scores (RPS) for probabilistic categorical predictions on the leave-one-out test datasets for the RESP and comparison models.  RPS scores are reported relative to the median RPS for the CLIM reference model's unconditional predictions.  The RESP model generally has better (i.e., lower) and slightly less variable skill than the ``Sub'' and ``Common'' comparison models.}\label{fig:crps_skill}
\end{figure}

\end{knitrout}

\section{Discussion}
\label{sec:discussion}

The RESP model \autoref{eq:stProcess} expands geostatistical frameworks that
incorporate the effect of both local and remote covariates on spatially
correlated responses, like precipitation, but can be extended to address
additional spatio-temporal modeling needs.  For example, while we use the RESP
model to draw inference on entire response fields, the model's
process-formulation also allows it to be applied to more standard
spatial interpolation problems as well.  Since there is great uncertainty
in global climate model (GCM) predictions of future precipitation, statistical
downscaling methods have been widely used in regional climate change studies.
Validating the RESP model on historical data marks an improvement on existing 
approaches and implies it can be used with GCM predictions of surface
temperatures and large-scale patterns to infer predictions for precipitation
from covariate data only. By comparison with the RESP model, other models
directly model less of the spatial structure in teleconnected data, but other
models have been studied in broader statistical contexts.  Fortunately, it is
possible to formulate the RESP model more broadly.

Many scientific disciplines work with spatially-referenced non-Gaussian data,
for which the RESP model can be adapted.  For example, the RESP model could be
adapted to study teleconnective effects on the number of large rain events,
which are important for many ecological systems and sectors of society.
Following approaches common to generalized linear models for spatial data, the
existing RESP response $Y\paren{\vs, t}$ may be reinterpreted as a latent
Gaussian field that helps parameterize the distribution for non-Gaussian
observations \citep{Diggle1998, Higgs2010}.  The primary technical
challenge for Bayesian implementations of such models is to develop efficient
estimation procedures since conjugacy is lost.

Modeling effects for multivariate remote covariates or data on large spatial
domains could both be facilitated by modeling spatial dependence with
sparse geostatistical models.  Inference and prediction for many geostatistical
models involves matrix operations with $O\paren{\ns^3}$  computational
complexity. Sparse geostatistical models can avoid these costs on large spatial
domains, for example, by using Gaussian Markov random field
approximations to specific classes of Gaussian fields with Mat\'{e}rn
covariances \citep{Lindgren2011a}, covariance tapering to generate
spatial covariance matrices with banded structure \citep{Furrer2006},
multiresolution covariance models \citep{Katzfuss2016}, or hierarchical
nearest neighbor models \citep{Datta2014c}.  While computational savings may
be minimal for small spatial domains like Colorado, they may offset
computational costs of estimating teleconnection effects for multiple sets of
remote covariates.  The RESP model may naturally be extended to include
multiple teleconnection effects \autoref{eq:teleconnectionTerm} to model
impacts from Pacific and Atlantic Ocean temperatures, for example.
Multivariate teleconnection effects can also be used to model impacts from
a vector $\v z\paren{\vr, t}\in\mathbb R^m$ of $m$ remote covariates at
location $\vr\in\Dz$.  Both extensions require sensibly modifying
the remote coefficient covariance function \autoref{eq:remote_cov_def} and will
yield likelihood structures similar to the RESP model \autoref{eq:stProcess},
especially if relationships between additional teleconnection effects are
modeled with separable covariances.

Non-stationary covariance models and temporal extensions can also allow the
RESP model to be applied to more diverse data and problems. While the
teleconnection term \autoref{eq:teleconnectionTerm} admits temporal
non-stationarity moderated by the remote covariates, modeling temporal
dependence across timepoints can allow the RESP model to be used in more
traditional forecasting problems.  Similarly, modeling spatial non-stationarity
can potentially improve model fit and prediction at unobserved spatial
locations.    In particular, nonstationary covariances could allow the remote
coefficients to vary temporally. This extension may be relevant because
\citet{Mason2001} find that teleconnection effects can vary across seasons.
As in \citet{Choi2015}, however, changes over time may be difficult to detect
because the effects tend to be weak.

Without considering any extensions, however, the RESP model yields additional
discussion about spatial modeling.  The RESP model's inclusion of dependence at
both long and short distances echoes descriptions of the screening effect
\citep{Stein2015}.  Carefully studying spectral densities of covariance
functions show that if they decay quickly enough, then spatial predictions are
primarily driven by data from nearby locations.  While the RESP model allows
distant locations to influence spatial prediction, the RESP model does not
contradict the screening effect because it explicitly models long range
dependence through the teleconnection term \autoref{eq:teleconnectionTerm} and
the screening effect is a property of local covariance functions
\autoref{eq:local_cov_def}.  Of similar subtlety, maps of estimated
teleconnection effects (\autoref{fig:teleconnection_estimates}) raise discussion
about uncertainty estimates for spatial patterns.  Significance in
\autoref{fig:teleconnection_estimates} is determined pointwise with respect to
the posterior distribution for $\alpha'\paren{\vs, 1}$ at each location so can
provide inference for teleconnection effects at individual points.  Here,
pointwise significance can help individual municipalities determine whether
they are strongly impacted by teleconnection effects and may benefit from
use of the RESP model.  Determining uncertainty for entire regions is a
multiple testing problem not considered in this study
\citep{Bolin2014, French2016}.  Uncertainties for entire regions are more
important, for example, when trying to estimate boundaries for polluted areas.

There is potential for more diverse application of the RESP model because
teleconnections exist in other fields, like ecology \citep{Brierley1999} and
human geography \citep{Seto2012}.  The model's general introduction
in \Cref{sec:development} as a spatial regression problem highlights a
less-common class of spatial analysis problems because it addresses problems
that require dependence at both long and short distances, at odds with typical
assumptions that data at distant points are effectively independent. While the
RESP model assumes the response and remote covariates are defined on disjoint
spatial domains, it suggests even broader classes of problems in which
overlapping domains characterize dependence between distant locations, or in
which teleconnected domains are not known a priori and need to be estimated.
The latter problem is reminiscent of general covariance or graphical model
structure estimation problems, which may provide possible directions for
future spatial statistics research topics.

\section*{Supplementary materials} Additional information and supporting
material for this article is available online at the journal's website.

\section*{Acknowledgements} We express our gratitude to
Michael Stein and Mikael Kuusela for discussions that helped enrich
interpretations of the RESP model. This material is based upon work supported
by the National Science Foundation under grant numbers AGS-1419558 and
DMS-1106862. Any opinions, findings, and conclusions or recommendations
expressed in this material are those of the authors and do not necessarily
reflect the views of the National Science Foundation.

\bibliographystyle{apacite}
\bibliography{references}

\end{document}

\singlespace

\maketitle

\doublespace

We provide key details used to derive several theoretical and
computational results as well as some supporting figures for the study of
Colorado precipitation in \Cref{sec:caseintro}.

\section{Linear algebra details}\label{linalgapp}

We briefly review key properties of Kronecker products that facilitate theoretical
and numerical computations involving the RESP model.

\subsection{Likelihood marginalization}\label{kronapp}

We note that $cA = A\otimes c$ for $c\in\mathbb R$ and $A\in\mathbb R^{m\times n}$.
This lets us use the mixed product rule for Kronecker products
\citep[Thm. 14.3]{Banerjee2014} to distribute matrix multiplication across
Kronecker products involving vectors.  For example, let $c\in\mathbb R^n$,
$A\in\mathbb R^{m\times n}$, and
$B\in\mathbb R^{n\times p}$, then
\begin{align*}
  \paren{A\otimes c} B = \paren{A\otimes c}\paren{B\otimes 1} =
  AB\otimes c .
\end{align*}

\subsection{Numerical evaluation of likelihood}\label{evalapp}

Evaluating the RESP likelihood involves computing matrix multiplications that
involve Kronecker products.  For example, let $A\in\mathbb R^{m\times n}$,
$B\in\mathbb R^{p\times q}$, and $C\in\mathbb R^{nq\times r}$.  While computing
$\paren{A\otimes B}C$ naively requires $O\paren{mnpqr}$ floating point operations,
it can be computed in $O\paren{mpqr + mnqr}$ floating point operations by recognizing
that
\begin{align*}
  \paren{A\otimes B}C =
    \begin{bmatrix}
      B \paren{\sum_{j=1}^n a_{1j} C_j} \\
      \vdots \\
      B \paren{\sum_{j=1}^n a_{mj} C_j} \\
    \end{bmatrix}
\end{align*}
where $C_j$ represents the $j^{th}$ $q\times r$ block matrix in $C$, i.e.,
that $C_j\in\mathbb R^{q\times r}$ for $j\in\brc{1,\dots,n}$ and
\begin{align*}
  C = \begin{bmatrix}
    C_1 \\
    \vdots \\
    C_n
  \end{bmatrix} .
\end{align*}

While \citet{Banerjee2014} discuss a similar idea in Section 14.7,
they limit their treatment to the case in which $C$ is a vector.  Their discussion
also does not present this direct form for numerical evaluation; they
present results that rely on the $\textrm{vec}\paren{\cdot}$ operation instead.

\section{Bayesian implementation of the RESP model}
\label{sec:detailed_implementation}

We adopt a hierarchical Bayesian framework and use a hybrid Gibbs sampler
for inference for the RESP model \autoref{eq:stProcess}
using the likelihood \autoref{eq:stPredProcessFull}.
The Bayesian framework allows estimates for the transformed teleconnection
effects \autoref{eq:reparam} to be computed directly from posterior samples
of $\tilde{\v\alpha}^*$ by using the definition for $\tilde{\v\alpha}'$
specified after \autoref{eq:basisSubstitution} to appropriately transform the
sampled teleconnection effects $\tilde{\v\alpha}^*$.

\subsection{Model likelihood}
\label{sec:likelihood}

Using Gaussian processes to specify the RESP model's \autoref{eq:stProcess}
randomness implies the data model is jointly normal for finite samples with
$\ns$ locations, $\nr$ remote locations, and $\nt$ time points.  Let the column
vectors $\v Y_t = \brk{ Y\paren{\v s_i, t}}_{i=1}^\ns \in \mathbb R^\ns$ and
$\v z_t = \brk{ z\paren{\v r_j, t}}_{j=1}^\nr \in \mathbb R^\nr$,
and the matrix
$X_t \in \mathbb R^{\ns\times p}$ with row vectors
$\v x\paren{\vs_i, t}^T$ for $i=1,\dots,\ns$
represent the observed response variables and covariates at time $t$; and let
the column vector
$\v\alpha(\v s) =
  \brk{ \alpha\paren{\v s, \v r_j} }_{j=1}^\nr \in\mathbb R^\nr$
represent the teleconnection coefficients for location $\v s$.
The reduced rank assumption lets us use the Kriging notation from
\autoref{eq:pred} to write $\v\alpha(\v s)=\v c^* \inv{{R^*}} \v\alpha^*(\v s)$.
The matrix $\v c^*\in\mathbb R^{\nr\times k}$ is built with row vectors
$\v c^*\paren{\v r_i}^T$ for $i=1,\dots,\nr$ that
contain the covariances among the teleconnection effect
$\alpha\paren{\vs, \vr_i}$ and the teleconnection effects at knot locations
\\ $\alpha\paren{\vs,\vr_j^*}, j=1,\dots,k$.
This yields the data model for
$\v Y = \brk{\v Y_{t_1}^T \dots \v Y_{t_\nt}^T}^T \in \mathbb R^{\ns\nt}$,
which is given by
\begin{align}
  \label{eq:stPredProcessFull}
    \given{ \v Y}
    {\v\beta, \tilde{\v\alpha}^*, R^*, \v c^*, \Sigma}
    \thicksim& \mathcal N
    \paren{ \v\mu_{\v Y},
      ~ I_\nt \otimes \Sigma
    }
\end{align}
in which
\begin{align*}
\v\mu_{\v Y} = \tilde{\v X} \paren{\v 1_\nt \otimes \v\beta} +
      \tilde{\v Z}^* \paren{\v 1_\nt \otimes \tilde{\v\alpha}^*},
\end{align*}
where $\otimes$ denotes the Kronecker product,
$\tilde{\v X} = \diag\brc{X_{t_1}, \dots, X_{t_\nt}}$,
\\ $\tilde{\v Z}^* = \diag\brc{I_\ns \otimes {\v z_{t_1}^*}^T, \dots,
  I_\ns \otimes {\v z_{t_\nt}^*}^T}$,
 ${\v z_t^*}^T = \v z_t^T \v c^* \inv{{R^*}} \in\mathbb R^{1\times k}$,
$\tilde{\v\alpha}^* = \brk{\v\alpha^*(\v s_i)}_{i=1}^\ns \in\mathbb R^{\ns k}$,
and $\Sigma \in\mathbb R^{\ns\times\ns}$ is the local covariance matrix with
entries $\Sigma_{ij} = C_w\brc{(\v s_i, t), (\v s_j, t)}$.  While the covariate
matrices $\tilde{\v X} \in\mathbb R^{\ns\nt\times p\nt}$ and
$\tilde{\v Z}\in\mathbb R^{\ns\nt\times\ns k\nt}$ are block diagonal,
we later introduce alternate notation
to make evaluating posterior distributions easier.

Each time-indexed block in the remote effects term
$\tilde{\v Z}^* \paren{\v 1_\nt \otimes \tilde{\v\alpha}^*}$ has the form
\\ $\paren{I_\ns \otimes {\v z_t^*}^T}\tilde{\v\alpha}^*$, which helps show
how each response $Y\paren{\vs,t}$ at time $t$ shares the same remote covariates
$\v z_t^*$.  Although the remote coefficients $\alpha\paren{\vs,\vr^*}$,
$\vs\in\Dy$ vary spatially across $\Dy$ for a fixed $\vr^*\in\Dz$, the RESP
model differs from traditional spatially varying coefficient models
\citep[Section 9.6]{Banerjee2015} because all of these remote coefficients
share the same covariate $z^*\paren{\vr^*, t}$.

The likelihood \autoref{eq:stPredProcessFull} changes subtly when reparameterizing
the teleconnection effects \autoref{eq:teleconnectionTerm} to interpret them
with respect to the spatial basis function transformation defined by
\autoref{eq:reparam}.  The spatial basis function expansion
\autoref{eq:remote_expansion} of the remote covariates $z\paren{\vr,t}$ yields
the substitution
\begin{align}
  \label{eq:basisSubstitution}
    \tilde{\v Z}^* \paren{\v 1_\nt \otimes \tilde{\v\alpha}^*} =
    \tilde{\v A} \paren{\v 1_\nt \otimes \tilde{\v\alpha}'}
\end{align}
in the likelihood \autoref{eq:stPredProcessFull}.  Where
$\tilde{\v A} =
  \diag\brc{I_\ns \otimes \v A_{t_1}^T, \dots, I_\ns \otimes \v A_{t_\nt}^T}$
and \\ $\tilde{\v\alpha}' =
  \paren{I_\ns \otimes W^T\v c^* \inv{{R^*}}} \tilde{\v\alpha}^*$, and
$\tilde{\v\alpha}^* \in\mathbb R^{\ns K}$ where the vector $\v A_{t}$ and
matrix $W$ form the matrix decomposition of the remote covariate vector $\v z_t$
when expanded by spatial basis functions  $\v z_t = W \v A_t$.  The column vector
$\v A_t = \brk{a_l\paren{t}}_{l=1,\dots,K}\in\mathbb R^K$ contains the weights
at time $t$ for the basis functions $\brc{\psi_l\paren{\vr}: l=1,\dots,K}$ ,
which are stored in the basis function matrix $W \in\mathbb R^{\nr\times K}$
with entries $W_{jl} = \psi_l\paren{\vr_j}$.

\subsection{Gibbs sampler}

We use conjugate prior distributions to specify our Bayesian model where
possible, setting $\v\beta \thicksim \mathcal N \paren{\v 0, \v\Lambda}$ for a
fixed prior covariance matrix $\v\Lambda$ and
$\sigma^2_w \thicksim \distn{IG}{a_{\sigma^2_w}, b_{\sigma^2_w}}$.
We use standard choices for weakly informative priors for the remaining
parameters:
$\sigma^2_\alpha \thicksim \distn{IG}{a_{\sigma^2_\alpha}, b_{\sigma^2_\alpha}}$,
$\sigma^2_\varepsilon \thicksim
  \distn{IG}{a_{\sigma^2_\varepsilon}, b_{\sigma^2_\varepsilon}}$,
$\rho_w \thicksim \distn{U}{a_{\rho_w}, b_{\rho_w}}$,
and
$\rho_\alpha \thicksim \distn{U}{a_{\rho_\alpha}, b_{\rho_\alpha}}$
\citep{Banerjee2008a}.
As Mat\'{e}rn smoothness parameters are difficult to estimate in standard
applications, we estimate $\nu_w$ and $\nu_\alpha$ from sample variograms and
treat these parameters as fixed during model fitting.

Bayesian estimation is often more stable after integrating out latent fields
\citep[pg. 126]{Banerjee2015}. The special case of Kronecker product rules
reviewed in Supplement A \Cref{kronapp} facilitates this integration.
The marginalized data likelihood with $\nt$ timepoints after
integrating out $\tilde{\v\alpha}^*$ is given by
\begin{equation}
  \given{\v Y}{\v\beta, R^*, \v c^*, \Sigma}
  \thicksim \mathcal N \paren{
    \tilde{\v X} \paren{\v 1_\nt \otimes \v\beta},
    ~ \inv C \otimes \Sigma
  }
\label{eq:stMarginalized}
\end{equation}
where $\inv C = I_\nt + {\v Z^*}^T R^* \v Z^*$.  Since the Kronecker
product is a bilinear operator, the marginalized variance
$\inv C \otimes \Sigma$ decomposes into the sum
$\paren{I_\nt \otimes \Sigma} + \paren{{\v Z^*}^T {{R^*}} \v Z^* \otimes \Sigma}$
which more clearly highlights how the remote covariates account for some of the
spatial variability around the fixed mean
$\tilde{\v X}\paren{\v 1_\nt \otimes \v\beta}$.

The data likelihood \autoref{eq:stMarginalized} is almost fully identified.
The spatial covariance matrix $\Sigma$ has entries
$\Sigma_{ij} = C_w\brc{(\v s_i, t), (\v s_j, t)}$ based on the covariance
function $C_w$ defined in \autoref{eq:local_cov_def}.  The covariance's scale
parameters $\sigma^2_w$ and $\sigma^2_\varepsilon$ are only identifiable with
respect to their product $\sigma^2_w\sigma^2_\varepsilon$. We use parameter
expansion to remedy the identifiability issue by reparameterizing the nugget
variance as $\sigma^2_\varepsilon = \sigma^2_w \tilde{\sigma}^2_\varepsilon$.
Therefore, in model fitting, we estimate $\tilde{\sigma}^2_\varepsilon$ instead
of estimating $\sigma^2_\varepsilon$ directly.

Model fitting employs a hybrid Gibbs sampler with adaptive random walk
Metropolis steps to estimate parameters for the marginalized likelihood
\autoref{eq:stMarginalized}.  Likelihood evaluation is aided by computations
detailed in Supplement A \Cref{evalapp} that efficiently
implement Kronecker product matrix multiplication.
Sampling proceeds by updating the regression coefficient and spatial variance
parameters $\v\beta$ and $\sigma^2_w$ by drawing from their full conditional
posterior distributions
\begin{eqnarray}
  \given{\v\beta}{\v\cdot} &\thicksim& \mathcal N
    \paren{
    \Sigma_{\given{\v\beta}{\v\cdot}}
    \v X^T
    \paren{C \otimes \inv\Sigma}
    \v Y,
    ~ \Sigma_{\given{\v\beta}{\v\cdot}}
    } ,
  \\
  \given{\sigma^2_w}{\v\cdot} &\thicksim& \distn{IG}{
    a_{\sigma^2_w} + \ns\nt / 2,
    ~ b_{\sigma^2_w} +
    e^T
    \brk{C \otimes \inv{\paren{\Sigma/\sigma^2_w}}}
    e
    / 2
  }
\end{eqnarray}
where $\v X \in \mathbb R^{\ns\nt \times p}$ is a block row matrix with
row blocks $X_{t_i}$ for $i=1,\dots,\nt$, the column vector
$e = \paren{\v Y - \tilde{\v X}\paren{\v 1_\nt \otimes \v\beta}}$
are the model residuals, and
\begin{equation}
\label{eq:postCovBeta}
  \Sigma_{\given{\v\beta}{\v\cdot}} =
    \inv{\brc{\inv{\v\Lambda} + \v X^T \paren{C \otimes \inv\Sigma} \v X}}
\end{equation}
is the posterior covariance matrix for $\v\beta$.  The remaining parameters
$\rho_w$, $\rho_\alpha$, $\sigma^2_\alpha$, and $\tilde\sigma^2_\varepsilon$ are
transformed to unconstrained supports and updated using adaptive random walk
Metropolis steps with normal proposals.  Log transformations are used for the
positive parameters $\sigma^2_\alpha$ and $\tilde\sigma^2_\varepsilon$, and logit
transformations are used for the bounded parameters $\rho_w$ and $\rho_\alpha$.
The adaptive proposal variance $\lambda\paren{\cdot}$ differs for each parameter and is
tuned at each iteration following a basic version of Algorithm 4 from \citet{Andrieu2008}.
Additional computations,
however, are required to estimate the remote coefficients and make predictions.

Composition sampling \citep[p. 126]{Banerjee2015}
provides a means to sample from the posterior distribution
of the remote coefficients $\tilde{\v\alpha}^*$
\autoref{eq:alpha_composition_posterior} as well as from the posterior
predictive distribution of the response variables $\v Y_{t_0}$ at a new
timepoint $t_0$.  This allows inference and prediction of these processes, which
the Gibbs sampler does not directly study.

The Gaussian process assumption and
separable covariance \autoref{eq:remote_cov_def} imply the remote coefficients
have prior distribution
\begin{equation}
  \given{\tilde{\v\alpha}^*}{\Sigma, R^*}
    \thicksim \mathcal N \paren{ \v 0, ~ \Sigma \otimes R^*} .
\end{equation}
The full conditional posterior distribution for $\tilde{\v\alpha}^*$ is
\begin{equation}
\label{eq:alpha_composition_posterior}
  \given{{\tilde{\v\alpha}^*}}{\v\cdot} \thicksim \mathcal N \paren{
    \v\mu_{\given{{\tilde{\v\alpha}^*}}{\v\cdot}},
    ~ \Sigma_{\given{{\tilde{\v\alpha}^*}}{\v\cdot}} } ,
\end{equation}
where ``$\v\cdot$'' represents conditioning on all remaining unknown quantities and
\begin{eqnarray*}
  \v\mu_{\given{{\tilde{\v\alpha}^*}}{\v\cdot}} &=&
    \sum_{t\in\mathcal T} \brc{
      \paren{\v Y_t - X_t \v\beta} \otimes
      \inv{\paren{ {\inv{ {R^{*}}}} + \v Z^* {{\v Z}^*}^T }} \v z_t^* } ,
  \\
  \Sigma_{\given{{\tilde{\v\alpha}^*}}{\v\cdot}} &=&
    \Sigma \otimes \inv{\paren{ {\inv{ {R^{*}}}} + \v Z^* {{\v Z}^*}^T }} ,
\end{eqnarray*}
and the matrix $\v Z^* \in\mathbb R^{k\times\nt}$ with column vectors
$\v z_{t_i}^*\in\mathbb R^{k}$ for $i=1,\dots,\nt$
is a dense matrix that contains the remote covariates.

\subsection{Computational approach for conducting inference on remote coefficients}
\label{sec:computing}

We use standard hierarchical Bayesian spatial modeling techniques to
draw inference on $\tilde{\v\alpha}^*$ through composition sampling
\citep[p. 126]{Banerjee2015}.  Composition sampling generates a posterior sample
$\brc{{\tilde{\v\alpha}^{*(1)}}, \dots, {\tilde{\v\alpha}^{*(G)}}}$ for
$\tilde{\v\alpha}^*$ by using the full conditional posterior distribution
\autoref{eq:alpha_composition_posterior} for $\tilde{\v\alpha}^*$ with a posterior
sample of the model parameters $\v\beta$, $\v\theta_w$, $\v\theta_\alpha$,
and $\sigma^2_\varepsilon$.  The composition samples may be drawn in parallel to
reduce the computation time because composition samples
${\tilde{\v\alpha}^{*(i)}}$ are independent given the posterior parameter samples.
Drawing inference on $\tilde{\v\alpha}^*$ also requires computational techniques
to reduce memory demands.

The composition sample for $\tilde{\v\alpha}^*$ requires
storing $\ns \times k \times G$ floating point numbers.  Even for moderately
sized studies with $\ns = 200$, $k = 30$, and $G=20,000$, the composition sample
requires 915MB of memory.  Although this demand increases linearly in
$k$, $\ns$, and $G$, it quickly becomes burdensome for typical personal computers.
We therefore estimate
 $\tilde{\v\alpha}^*$ using the normal approximation to the posterior.
The normal approximation only requires the composition sample's mean
$\hat{\v\mu}_{\given{ {\tilde{\v\alpha} } }{\v Y}} =
  \frac{1}{G}\sum_{g=1}^G { \tilde{\v\alpha}^{*(g)} }$
and covariance matrix
$\hat{\Sigma}_{\given{ {\tilde{\v\alpha}} }{\v Y}} =
  \frac{1}{G-1}\sum_{g=1}^G
  \paren{\tilde{\v\alpha}^{*(g)} - \hat{\v\mu}_{\given{ {\tilde{\v\alpha} } }{\v Y}}}
  \paren{\tilde{\v\alpha}^{*(g)} - \hat{\v\mu}_{\given{ {\tilde{\v\alpha} } }{\v Y}}}^T$.
These objects require storing $\paren{\ns\times k}\paren{\ns\times k + 3}/2$
floating point numbers, which can dramatically reduce memory requirements
when $G>\paren{\ns\times k + 3}/2$.

We use strategies from \citet{Pebay2008} to facilitate computing
these summary objects with minimal memory requirements.
We use partitions of the composition samples and compute
$\hat{\v\mu}_{\given{ {\tilde{\v\alpha} } }{\v Y}}$ and
$\hat{\Sigma}_{\given{ {\tilde{\v\alpha}} }{\v Y}}$ in a streaming fashion, which
allows estimation of $p\paren{\given{\tilde{\v\alpha}^*}{\v Y}}$
in parallel and with minimal memory requirements
\citep[eqs. 1.1, 1.3, 3.11, \& 3.12]{Pebay2008}.
These benefits are achieved by recognizing, for example, that a running estimate of
$\hat{\v\mu}_{\given{ {\tilde{\v\alpha} } }{\v Y}}$ based on
$\brc{{\tilde{\v\alpha}^{*(1)}}, \dots, {\tilde{\v\alpha}^{*(g)}}}$ is easy to
update when the next composition sample ${\tilde{\v\alpha}^{*(g+1)}}$
is drawn, and that the updating equations yield
$\hat{\v\mu}_{\given{ {\tilde{\v\alpha} } }{\v Y}}$ after all $G$ composition
samples are processed.

\section{Implementation of model assessment measures}
\label{sec:assessment}

\subsection{Variance inflation factors for local effects}

The posterior covariance matrix \autoref{eq:postCovBeta}
for the regression coefficient vector $\v\beta$
allows us to follow \citet{Reich2006} and define conditional variance
inflation factors that can help diagnose multicollinearity
between the local and remote covariate matrices $\v X$ and $\v Z$ via
\begin{equation*}
  \textrm{VIF}\paren{\beta_i} =
    \frac{\paren{\inv{\brc{
      \inv{\v\Lambda} +
      \v X^T \paren{C \otimes \inv\Sigma} \v X
    }}}_{ii}}
    {\paren{\inv{\brc{
      \inv{\v\Lambda} +
      \v X^T \paren{I_\nt \otimes \inv\Sigma} \v X
    }}}_{ii}} ~.
\end{equation*}
The VIF measures the proportional increase in the $i^{th}$ local coefficient's
posterior variance caused by adding remote covariates to the model, conditional
on the model's covariance parameters.
This interpretation follows since the denominator represents the
$i^{th}$ local coefficient's posterior covariance in a standard spatial
regression model where the local covariates and responses are observed at
multiple, independent timepoints. Larger VIF values indicate greater
multicollinearity, while the smallest possible VIF value of 1 indicates
no multicollinearity.  The VIF for $\beta_T$ is 1.1, which indicates that
estimates of the local effects are not impacted by the addition of remote
covariates in the case study of Colorado winter precipitation
(\Cref{sec:caseintro}).

\subsection{Variance inflation factors for remote effects}

As with the local effects, we can use the posterior covariance matrix in
\autoref{eq:alpha_composition_posterior} for the remote effects vector
$\tilde{\v\alpha}^*$ to define conditional variance inflation factors that can
help diagnose multicollinearity in teleconnection effects
$\brc{\alpha\paren{\vs,\vr^*_i}:\vs\in\Dy}$ associated with the $i^{th}$ knot
location $\vr^*_i$ denoted by
\begin{equation*}
  \textrm{VIF}\paren{\vr^*_i} =
    \frac{\paren{\inv{\paren{ {\inv{ {R^{*}}}} + \v Z^* {{\v Z}^*}^T }}}_{ii}}
    {\inv{\paren{
      1/\sigma^2_\alpha + \v Z^*_{i,\cdot} {{\v Z}^*_{i,\cdot}}^T
    }}} ~.
\end{equation*}
The notation $\v Z^*_{i,\cdot}$ indicates the $i^{th}$ row of the matrix
$\v Z^*$, in which row $i$ contains all observations of the remote covariate
at location $\vr^*_i$,  should be used in computations.
Conditional on the model's covariance
parameters, the VIF measures the proportional increase in the marginal posterior
variance of teleconnection effects $\brc{\alpha\paren{\vs,\vr^*_i}:\vs\in\Dy}$ associated
with the $i^{th}$ knot location $\vr^*_i$ that results from adding the remaining
remote covariates $\brc{z\paren{\vr_j, t} : j\neq i, ~ t\in\mathcal T}$ to
the model.  In the case study of Colorado winter precipitation (\Cref{sec:caseintro}),
$\textrm{VIF}\paren{\vr^*_i}$ ranged between 1.0 and 1.1 for all knot locations
$\vr^*_i$.  This indicates the reduced rank approximation and choice of knot
locations mitigates potential multicollinearity in estimation of teleconnection
effects.

\subsection{Heidke skill score}
\label{sec:hs}

The Heidke skill score ($\HS$) evaluates categorical predictions and is
commonly used in climate science \citep[Section 18.1]{VonStorch1999}. The measure
compares the probability the RESP model correctly predicts
precipitation $p_{\textrm{RESP}}$ to the probability that a reference model
correctly predicts precipitation $p_{\textrm{Ref}}$.
We adopt a standard, naive reference model that assigns equal probability to
all precipitation levels, implying $p_{\textrm{Ref}}=1/3$ since our precipitation
levels represent empirical terciles.  We additionally
manipulate the Heidke skill score formula \citep[eq. 18.1]{VonStorch1999}
to show that it is linear in $p_{\textrm{RESP}}$.  The manipulation also
yields an intuitive interpretation of the score:
\begin{align*}
\HS\paren{\textrm{RESP}} =
  \frac{p_{\textrm{RESP}} - p_{\textrm{Ref}}}{1 - p_{\textrm{Ref}}} =
  \paren{p_{\textrm{RESP}}/p_{\textrm{Ref}}-1} \Odds{p_{\textrm{Ref}}} .
\end{align*}
The Heidke skill score scales the odds that the reference model correctly predicts
precipitation by the RESP model's relative change in prediction accuracy.
Models have positive Heidke skill when they are more accurate than
the reference model; the maximum possible score is 1.
Similarly, models have negative skill when they are less accurate than the
reference model.

The predictive distributions for the
RESP, ENSO-T, and CLIM models are continuous, but easily discretized with
respect to the category cutpoints that are empirically determined from the
leave-one-out training data.  Therefore, the posterior mode of these distributions
provides a natural choice for categorical point predictions of precipitation.
The CCA model only produces continuous point predictions, so its categorical predictions
are defined by the category that matches the “above average”, “near average”,
or “below average” range in which the continuous point prediction lies.

The RESP model \autoref{eq:stProcess} frequently yields better point predictions
than the comparison models (\autoref{fig:hs_skill}). The CCA Heidke skill scores
are lower, but about as variable as the RESP model, which suggests the CCA model
may adequately account for spatial dependence when making predictions, but loses
skill by not also incorporating local covariates.  Similarly, the ENSO-T model's
Heidke skill scores may be more variable than the RESP model's scores since it
does not account for spatial dependence.

\begin{knitrout}
\definecolor{shadecolor}{rgb}{0.969, 0.969, 0.969}\color{fgcolor}\begin{figure}

{\centering \includegraphics[width=\maxwidth]{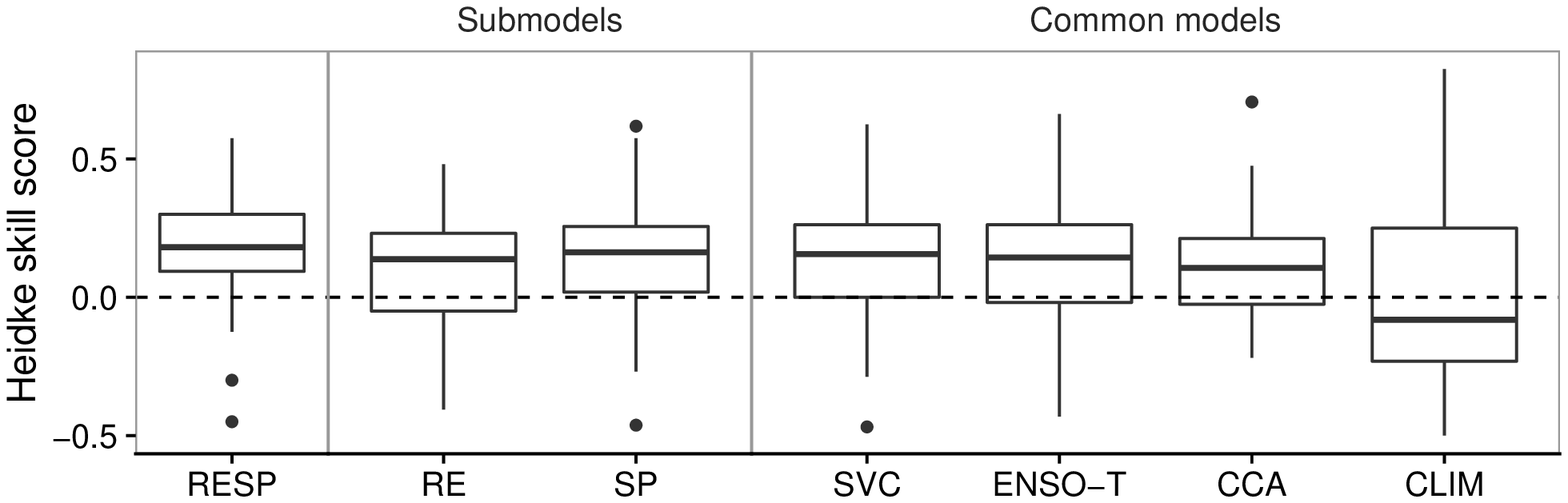} 

}

\caption[Comparison of Heidke skill scores for categorical point predictions on the leave-one-out test datasets for the RESP, ENSO-T, CCA, and CLIM models]{Comparison of Heidke skill scores for categorical point predictions on the leave-one-out test datasets for the RESP, ENSO-T, CCA, and CLIM models.  The dashed line at 0 marks the Heidke skill for a naive reference model that produces random point predictions  The RESP model generally has better (i.e., higher) and less variable skill than the comparison models.}\label{fig:hs_skill}
\end{figure}

\end{knitrout}

\subsection{Ranked probability score}

The ranked probability score (RPS) is closely related to the continuous ranked
probability score (CRPS), which is a proper scoring rule for probability measures
with finite means \citep{Gneiting2007a}.  \citet{Epstein1969} introduced the RPS
to evaluate probabilistic predictions of ordinal variables.  \citet{Murphy1971}
presented an equivalent formulation of the RPS that shows how the RPS, like the
CRPS, sums the squared differences between the predicted and observed cumulative
distribution functions for an ordinal response $Y_o\paren{\vs,t}$ at location
$\vs$ and time $t$.  Since the RPS is defined pointwise, we follow common
practice in climatological applications and average RPS scores over $\ns$
locations at which we observe the response $Y_o\paren{\vs,t}$ \citep{Hersbach2000},
yielding the RPS score at time $t$ for a model $\mathcal M$ that predicts the
cumulative distribution $\hat F_t\paren{j; \vs}$ for $j\in\brc{1,\dots,k}$
ordered categories of the response at location $\vs$
\begin{equation*}
\textrm{RPS}\paren{\mathcal M, t} = \frac{1}{\ns}\sum_{i=1}^\ns \sum_{j=1}^k
  \paren{\hat F_t\paren{j; \vs_i} - \mathds{1}\brc{Y_o\paren{\vs_i,t} \leq j}} ^2.
\end{equation*}

\section{Model validation maps}\label{predmapapp}

\Cref{sec:predresults} builds support for the RESP model by comparing it
to submodels and alternatives (\autoref{fig:crps_skill}).
\autoref{fig:fcst_std_errs} and \autoref{fig:cat_forecast_plots} show
continuous and discretized (categorical) predictions and uncertainties for the
1982 validation set.  Shrinkage and uncertainty in the
continuous predictions can be anticipated because even though teleconnection
effects contain predictive information, their overall influence on precipitation
tends to be relatively weak (\autoref{fig:tele_eda} B).

The categorical predictions for average monthly precipitation in winter across
Colorado are better determined near teleconnected regions.  Posterior logits
(\autoref{fig:cat_forecast_plots} C) for the categorical forecasts
(\autoref{fig:cat_forecast_plots} B) quantify uncertainty on an interpretable
scale in this application.  Since the modes of the discretized posterior
predictive distributions are the categorical forecasts, odds compare the
the forecasted category probabilities to the other categories.
The logit (log-odds) allows zero to be a natural reference point for comparing
uncertainties. Since we discretize posterior predictive distributions into
three categories, the probability for each categorical forecast is at least 1/3.
Non-negative logits indicate at least 50\% greater certainty since their
categorical forecast probability is at least 1/2.  Regions with non-negative
logits occur near locations with significant teleconnection effects
(\autoref{fig:teleconnection_estimates}).

\begin{knitrout}
\definecolor{shadecolor}{rgb}{0.969, 0.969, 0.969}\color{fgcolor}\begin{figure}

{\centering \includegraphics[width=\maxwidth]{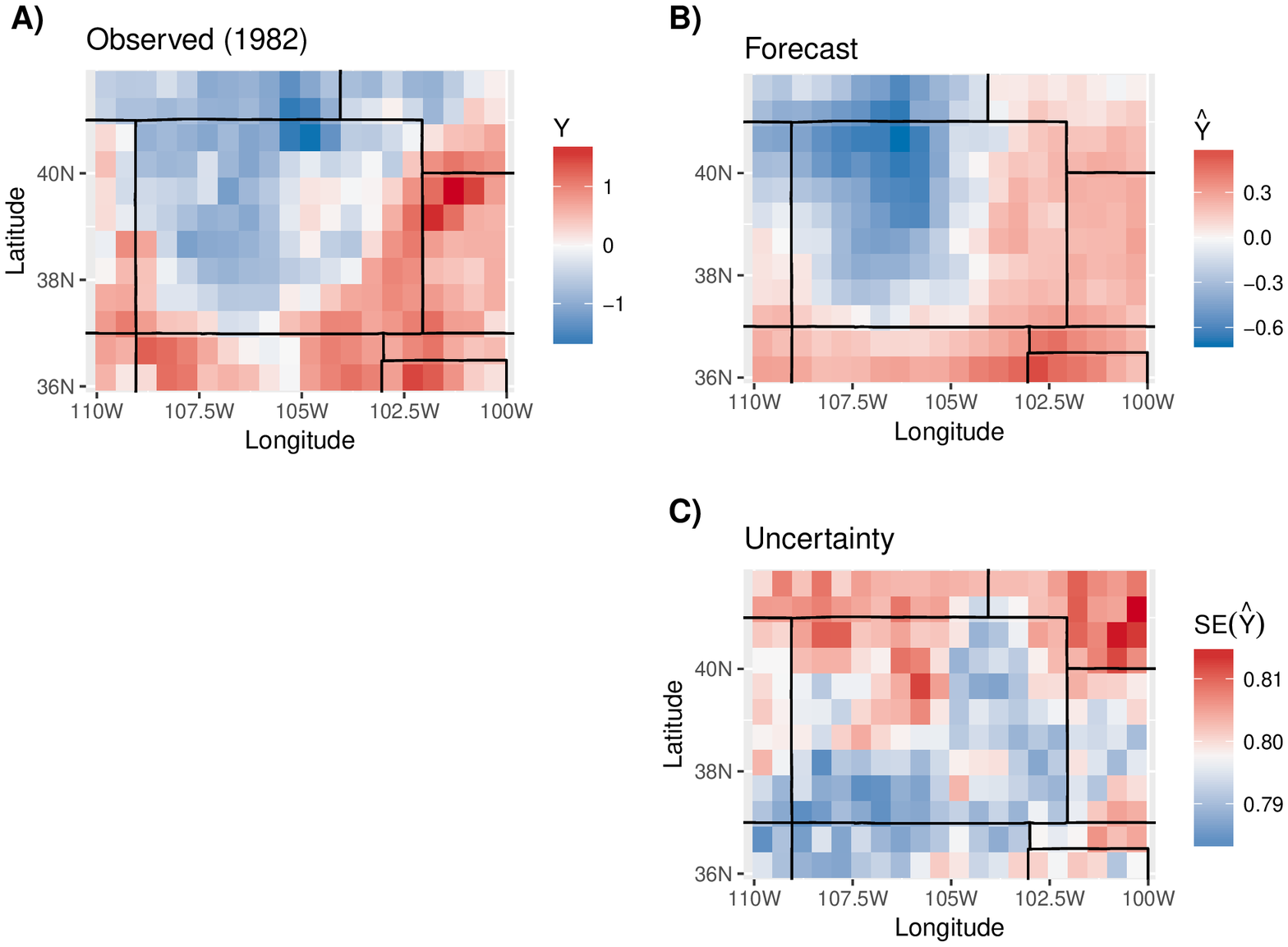} 

}

\caption[Comparison of predictions for the 1982 validation set in the leave-one-out cross-validation study]{Comparison of predictions for the 1982 validation set in the leave-one-out cross-validation study.  The pattern of the posterior predictive means (B) matches the PRISM responses (A) well, but the color scale indicates shrinkage of the forecasted magnitudes.  Posterior predictive standard errors (C) indicate uncertainty for the forecast.}\label{fig:fcst_std_errs}
\end{figure}

\end{knitrout}

\begin{knitrout}
\definecolor{shadecolor}{rgb}{0.969, 0.969, 0.969}\color{fgcolor}\begin{figure}

{\centering \includegraphics[width=\maxwidth]{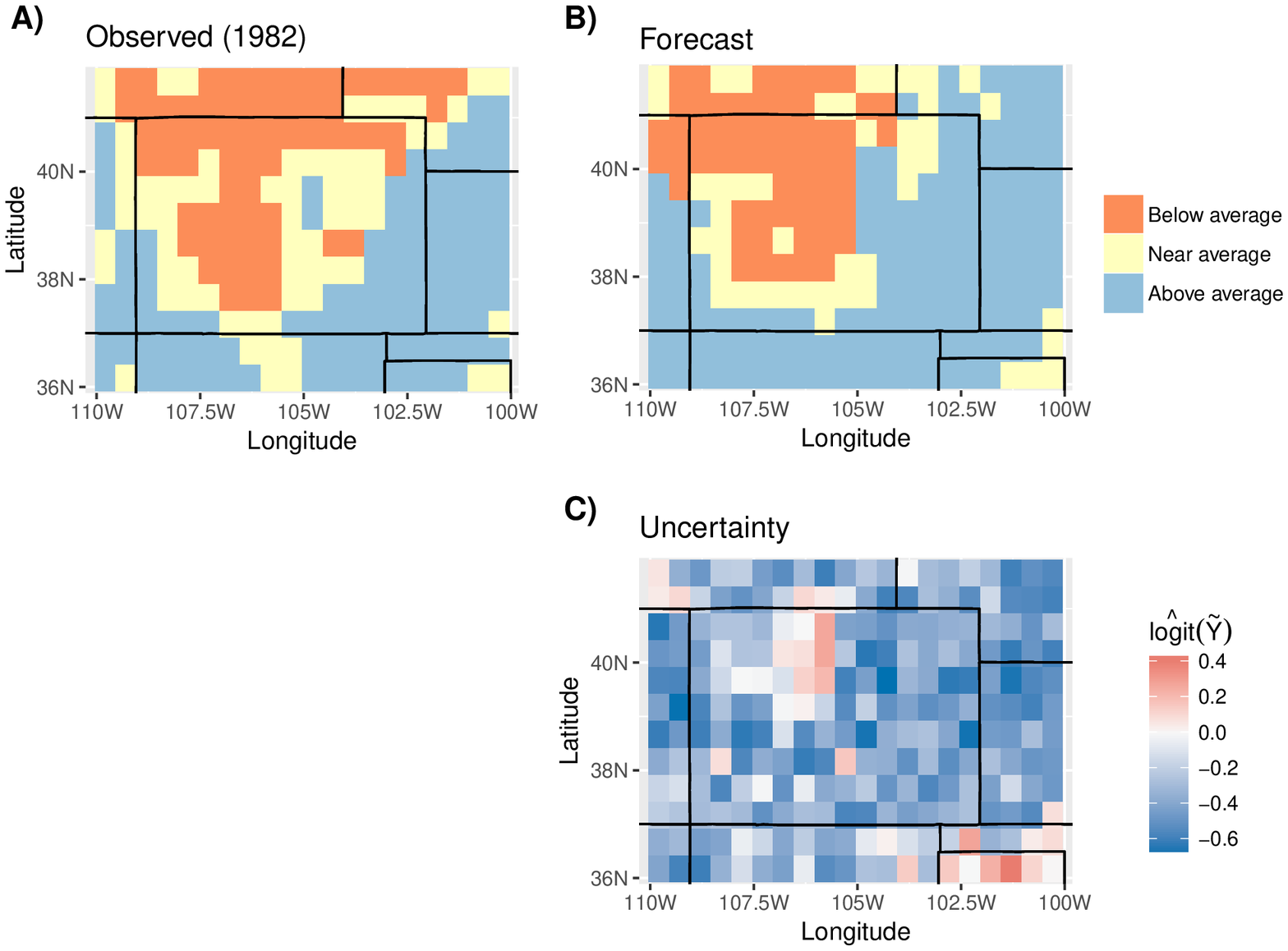} 

}

\caption[Comparison of discretized (categorical) predictions for the 1982 validation set in the leave-one-out cross-validation study]{Comparison of discretized (categorical) predictions for the 1982 validation set in the leave-one-out cross-validation study.  The pattern of the posterior predictive modes (B) matches the PRISM responses (A) well.  Logits for the posterior predictive modes (C) indicate higher uncertainty for the forecasts (blue), especially in regions without significant teleconnection effects.}\label{fig:cat_forecast_plots}
\end{figure}

\end{knitrout}

\clearpage

\begin{knitrout}
\definecolor{shadecolor}{rgb}{0.969, 0.969, 0.969}\color{fgcolor}\begin{figure}[b]

{\centering \includegraphics[width=\maxwidth]{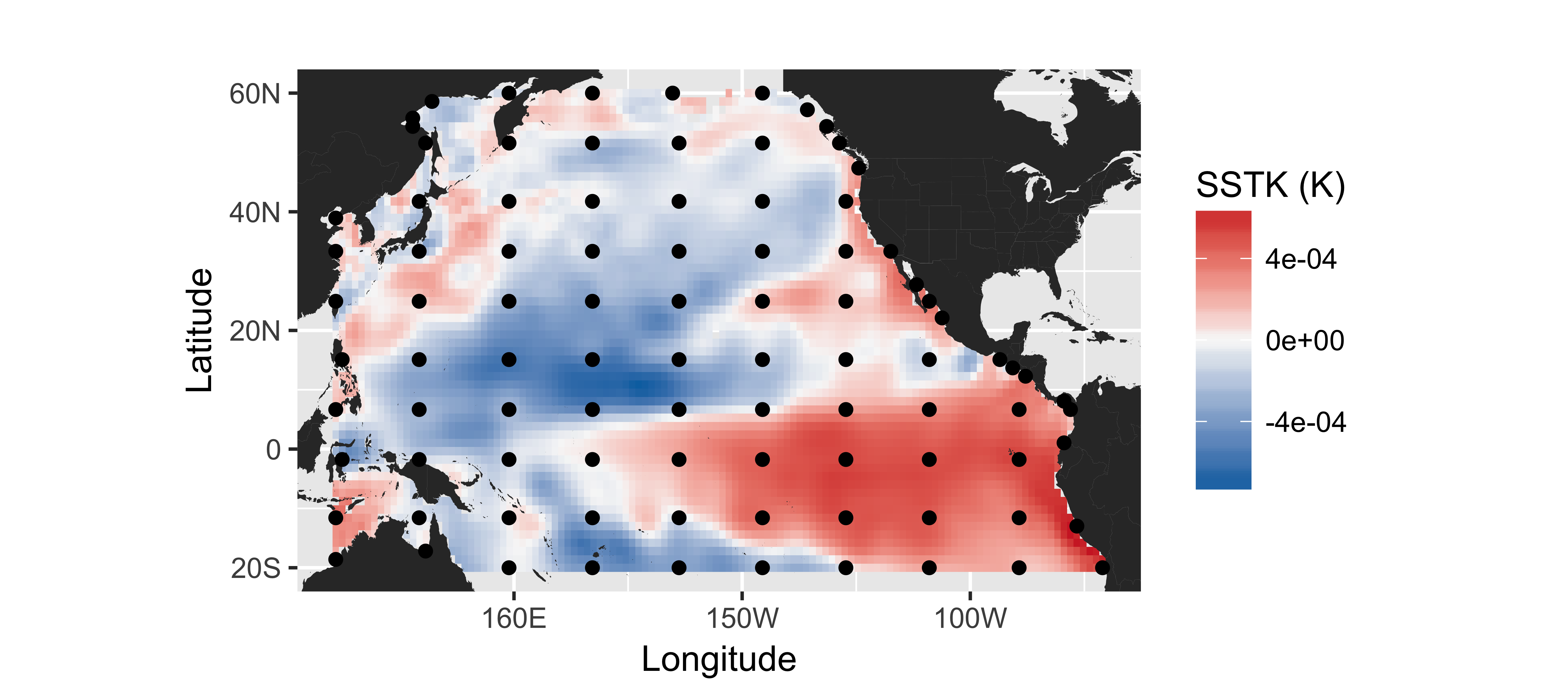} 

}

\caption[Average monthly  sea surface temperature standardized anomalies from Winter, 1982]{Average monthly  sea surface temperature standardized anomalies from Winter, 1982.  Black dots mark knot locations.}\label{fig:knot_locations}
\end{figure}

\end{knitrout}

\FloatBarrier

\bibliographystyle{apacite}
\bibliography{../references}